\documentclass[11pt,onecolumn]{IEEEtran}
\usepackage{graphicx}
\usepackage{amsmath}
\usepackage{amssymb}
\usepackage{setspace}
\usepackage{cite}
\usepackage[printonlyused]{acronym}
\usepackage{cite}
\usepackage{bm,multirow, rotating}
\usepackage{epstopdf}
\usepackage{rotating}
\usepackage{pdflscape}
\usepackage{subcaption}
\usepackage{hyperref}
\usepackage[dvipsnames]{xcolor}
\colorlet{hai}{black}

\begin{document}
\title{Spatial Reverberation and Dereverberation using an Acoustic Multiple-Input Multiple-Output System}
\author{Hai~Morgenstern*~and~Boaz~Rafaely 
\thanks{*The authors are with the Department of Electrical and Computer Engineering, Ben-Gurion University of the Negev, Be'er-Sheva 84105, Israel (email: \{haimorg\}@post.bgu.ac.il)}}

\maketitle
\begin{abstract}
Methods are proposed for modifying the reverberation characteristics of sound fields in rooms by employing a loudspeaker with adjustable directivity, realized with a compact spherical loudspeaker array (SLA).
These methods are based on minimization and maximization of clarity and direct­-to-­reverberant sound ratio. 
Significant modification of reverberation is achieved by these methods, as shown in simulation studies.
The system under investigation includes a spherical microphone array and an SLA, comprising a multiple-­input multiple-­output system.
The robustness of these methods to system identification errors is also investigated. 
Finally, reverberation and dereverberation results are validated by a listening experiment.
\end{abstract}


\section{INTRODUCTION}

Room reverberation has been shown to have both desired and undesired effects on the perception of sound by human listeners and on the performance of acoustic systems and applications. 
For example, while a long reverberation time (RT) contributes to listening qualities such as warmth, brilliance, and fullness of tone \cite{beranek2012concert}, it can degrade speech intelligibility and the performance of speech recognition and source localization systems \cite{bistafa2000reverberation,kinoshita2013reverb,gillespie2002acoustic,gustafsson2003source}. 
Methods have been proposed both for increasing reverberation, and for reducing it (dereverberation), using information about the room impulse response (RIR) between a loudspeaker and a microphone. 
In acoustic channel equalization, for example, an inverse system design is used to equalize the effects of the RIR on the input signal. 
However, inverting an RIR is challenging in practice \cite{neely1979invertibility}. 
To avoid system inversion, room-reverberation compensation (RRC), or listening room compensation, has been proposed \cite{kallinger2005impulse,mertins2010room}. 
RRC takes into account psychoacoustic measures, and an RIR is equalized only partially so as to remove \textit{audible} reverberation. 
Such psychoacoustic measures include, for example, the temporal masking curve of the human auditory system and the clarity index, $C50$ \cite{standard19973382}.
In particular, equalization filters have been designed for maximizing $C50$ \cite{kallinger2005impulse}. 
These methods were then generalized to apply to multichannel systems with multiple loudspeakers, multiple microphones, or both. 
Acoustic channel equalization was extended to the mutliple-input/output inverse theorem (MINT) \cite{miyoshi1988inverse}, providing an inverse filter under some restricting assumptions. 
Similarly, RRC was also extended to apply to multichannel systems in \cite{zhang2010use}, and methods were proposed for increasing robustness against RIR estimation errors \cite{lim2014robust,mei2010robustness,jungmann2012combined}.
In particular, in \cite{mei2010robustness} an RIR-shaping method, in which the magnitude of individual reflections in the RIR is controlled, was proposed.  

The multichannel input methods presented above typically position the loudspeakers at distinctively different locations to increase the spatial diversity of the system. 
While this approach is effective for improving equalization, it leads to a non-collocated source configuration.  
Recent studies demonstrated the benefits of using collocated, or compact, acoustic sources  \cite{pollow2009variable,warusfel2001directivity,rafaely2009spherical,morgenstern2015theory}.
In addition to the practical benefit of having a compact system, the directivity of the compact source can be controlled to simulate real sources (e.g., musical instruments), by synthesizing the sound field generated by these instruments \cite{warusfel2001directivity}. 
In particular, compact spherical loudspeaker arrays (SLAs) have been studied recently, and were shown to have the following desirable properties: (i) they can efficiently radiate sound in all directions, and (ii) they can produce complex radiation patterns despite their compactness \cite{kassakian2003design,zotter2007modeling}.
The employment of such arrays in room acoustic measurements has been proposed and implemented in \cite{gerzon1975recording} and \cite{farina2006room}, respectively, and a framework for spatially analyzing sound fields produced by SLAs has been presented in \cite{morgenstern2015theory}. 
The use of SLAs in sound field analysis has been illustrated in \cite{morgenstern2015theory,morgenstern2012joint,morgenstern2013enhanced}, and they have also been studied for sound field synthesis, for active noise cancellation, and for synthesizing radiation patterns of musical instruments \cite{pasqual2010application,pollow2009variable,rafaely2009spherical}. 
However, the potential of compact SLAs for modifying the reverberation characteristics of sound fields in rooms has not been explored extensively.

Recently, a method for both dereverberation and reverberation has been proposed, that controls the sound field at a region in a room by applying beamforming to an SLA \cite{morgenstern2013sound}. 
In particular, it has been shown that sound field characteristics, such as the level of reverberation and the strength of individual reflections, can be manipulated. 
However, this was only a preliminary study, showing limited results for a specific case of SLA and room parameters.  
This paper extends the results of \cite{morgenstern2013sound}, with the following additional contributions: 
\begin{enumerate}
\item[i] the formulation of a more comprehensive system model; an analytical description of the point-to-point transmission between the SLA and a listener position is given as a function of the beamforming coefficients at the SLA. 
The relation between the directivity patterns of the proposed beamformers and the sound field produced at the listener position is also clearly outlined. 
\item[ii] an extended simulation study; the study shows that significant modification of reverberation can be achieved and offers an analysis of the robustness of the proposed beamformers. 
\item[iii] a formal listening test; the test validates the simulation results.
\end{enumerate}
The paper shows that both the spatial and the \textcolor{hai}{spectro-temporal} attributes of room reverberation at the position of a listener can be significantly modified by controlling the directivity of a compact SLA in a reverberant room to minimize or maximize $C50$ or the direct-to-reverberant ratio (DRR).

This paper is organized as follows.
In Sec.~\ref{sec:model}, a model is presented for a MIMO system positioned in a room, which comprises an SLA and a spherical microphone array (SMA). 
This model facilitates the derivation of SLA beamforming methods for dereverberation and reverberation, which are outlined in Sec.~\ref{sec:methods}. 
The characteristics of the sound field produced using these methods are investigated in an extensive simulation study in Sec.~\ref{sec:simulationstudies} for several room-acoustic and SLA parameters. 
Since the developed methods assume perfect knowledge of the RIRs between the SLA and the SMA, this section also provides an analysis of the robustness to RIR estimation errors. 
The paper concludes, in Sec.~\ref{sec:listeningtests}, with a subjective evaluation of the methods, using listening tests conducted with binaural sound reproduction and a head tracking system.   

\section{SYSTEM MODEL}\label{sec:model}

A model is presented for an acoustic MIMO system that comprises an SLA with $L$ loudspeakers mounted on a rigid sphere with radius $r_L$, and an SMA with $M$ microphones distributed over a rigid sphere with radius $r_M$. 
\textcolor{hai}{In particular, the system model formulation includes a normalization scheme that effectively removes the acoustic effects of the physical construction of the SLA and the SMA. 
This is intended to enhance their spatial resolution \cite{rafaely2011optimal,park2005sound}.}  
For the complete formulation of this model, and a study of its properties, the reader is referred to \cite{morgenstern2015theory}.

\begin{figure}
\centering
\includegraphics[trim = 5mm 90mm 30mm 40mm, clip, width=23pc]{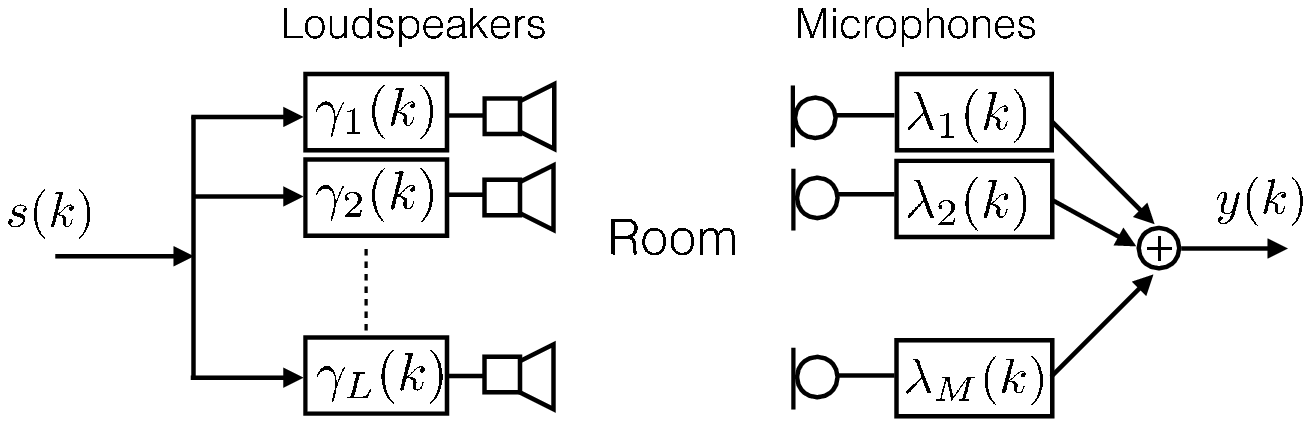}
\caption{System block diagram with SLA beamforming coefficients, $\gamma_1(k), ..,\gamma_L(k)$, and SMA beamforming coefficients, $\lambda_1(k),..., \lambda_M(k)$.}
\label{fig01}
\end{figure}

Beamforming is applied to both arrays with the aim of controlling the arrays' directivity patterns. 
At the SLA, this is achieved by weighting the scalar input signal, $s(k)$, where $k$ is the wavenumber, with beamforming coefficients $\gamma_1(k), ...,\gamma_L(k)$, before driving the loudspeakers. 
At the SMA, the microphone signals are weighted with beamforming coefficients $\lambda_1(k),..., \lambda_M(k)$ and then summed to produce the scalar system output, $y(k)$.
This is formulated as: 
\begin{eqnarray}\label{eq:bbase0}
y(k)& =&    \tilde{\bm \gamma}(k)^{\mathrm{H}}   \tilde{ \bm H}(k) \tilde{\bm \lambda}(k)   s(k), 
\end{eqnarray} 
where 
\begin{eqnarray}\label{eq:bbase1}
\tilde{\bm \gamma}(k) & =&    [\gamma_1(k),\, \gamma_2(k),...,\,\gamma_L(k)]^\mathrm{T}, \\
\tilde{\bm \lambda}(k) & =&    [\lambda_1(k), \,\lambda_2(k),...,\,\lambda_M(k)]^\mathrm{T},  \label{eq:bbase15}
\end{eqnarray} 
and $(\cdot)^\mathrm{H}$ and $(\cdot)^{\mathrm T}$ are the conjugate transpose operator and the transpose operator, respectively.  
Finally, $\tilde{ \bm H}(k)$ is a $L\times M$ matrix whose elements hold the room transfer functions (RTFs) between all loudspeaker and microphone combinations at wavenumber $k$. 
For a system block diagram representing Eq.~(\ref{eq:bbase0}), the reader is referred to Fig.~\ref{fig01}.

Next, the system in Eq.~(\ref{eq:bbase0}) is reformulated in the spherical harmonics (SHs) domain. 
Thus, RTFs are formulated between $(N_L+1)^2\leq L$ SHs channels of the SLA and $(N_M+1)^2\leq M$ SHs channels of the SMA \cite{morgenstern2015theory}, instead of between the $L$ loudspeakers and the $M$ microphones, respectively.
An SLA SHs channel is denoted by an order, $n\leq N_L$, and degree, $-n \leq m \leq n$. 
Similarly for the SMA, a SHs channel is denoted using order $n'\leq N_M$ and degree $-n'\leq m' \leq n'$.
Beamforming can now be applied to both of the arrays directly in the SHs domain, as a weighted summation of the arrays' SHs channels;  
Beamforming coefficients, $\gamma_{nm}(k)$, are applied to the SLA SHs channels instead of $\gamma_l(k)$ and, similarly, $\lambda_{n'm'}(k)$ are applied to the SMA SHs channels. 
The system output can now be reformulated using: 
\begin{eqnarray}\label{eq:bcase0}
y(k)& =&    \bar{\bm \gamma}(k)^{\mathrm{H}}   { \bm H}(k) \bar{\bm \lambda}(k)   s(k), 
\end{eqnarray} 
where 
\begin{eqnarray}\label{eq:bbase1b}
\bar{\bm \gamma}(k) & =&    [\gamma_{00}(k),\, \gamma_{1(-1)}(k),\, \gamma_{10}(k),\, \gamma_{11}(k),...,\,\gamma_{N_LN_L}(k)]^\mathrm{T}, \\
\bar{\bm \lambda}(k) & =&     [\lambda_{00}(k),\, \lambda_{1(-1)}(k),\, \lambda_{10}(k),\, \lambda_{11}(k),...,\,\lambda_{N_MN_M}(k)]^\mathrm{T},
\end{eqnarray} \label{eq:bbase15b}
and ${ \bm H}(k)$ is a $(N_L+1)^2\times (N_M+1)^2$ matrix whose elements hold the RTFs between all SLA and SMA SHs combinations. 
In particular, for free-field conditions a simplified far-field model can be used to provide a closed form analytical description of the sound field \cite{morgenstern2014farfield}. 
In this case, ${ \bm H}(k)$ can be written as \cite{morgenstern2015theory}:
\begin{eqnarray}\label{eq:bbase1c}
\bm H(k)& =&  \bm h^L_0(k) [\bm h^M_0(k)]^\mathrm{H}, 
\end{eqnarray} 
\textcolor{hai}{where
\begin{eqnarray}
\bm h^L_0(k) & = & \begin{bmatrix}  [h^L_{0}(k)]_{00},\, [h^L_{0}(k)]_{1(-1)},\, [h^L_{0}(k)]_{10},\, [h^L_{0}(k)]_{11},...,\,[h^L_{0}(k)]_{N_LN_L} \end{bmatrix}^\mathrm{T} \,\, \text{and}  \\
\bm h^M_0(k) & = & \begin{bmatrix}  [h^M_{0}(k)]_{00},\, [h^M_{0}(k)]_{1(-1)},\, [h^M_{0}(k)]_{10},\, [h^M_{0}(k)]_{11},...,\,[h^L_{0}(k)]_{N_MN_M} \end{bmatrix}^\mathrm{T} 
\end{eqnarray}
are the $(N_L+1)^2 \times 1$ SLA and $(N_M+1)^2 \times 1$ SMA RTF vectors, respectively. 
The elements of $\bm h^L_0(k)$ and $\bm h^M_0(k)$ are given by: 
\begin{eqnarray}\label{eq:bbase01}
[ h^L_0(k)]_{nm} & = & \frac{ e^{ikr_0}}{r_0} b_{n}^{L}(kr_L) Y_{n}^{m}(\bm \beta_0), \, \text{and}
\end{eqnarray}
\begin{eqnarray}\label{eq:bbase02}
[ h^M_0(k)]_{n'm'} & =  & [b_{n'}^{M}(kr_M)]^* Y_{n'}^{m'}(\bm \xi_0),  
\end{eqnarray} 
respectively.
In these equations, 
\begin{eqnarray}\label{eq:shy1}
b_{n}^L(kr_L) & = &  \rho c r^2_L (-i)^{n+1}\left( j_{n}(kr_L) - \frac{j'_{n}(kr_L)}{h'_{n}(kr_L)}h_{n}(kr_L) \right)\,\,\, \text{and}\\
b_{n'}^M(kr_M) & = & 4\pi(-i)^{n'}\left( j_{n'}(kr_M) - \frac{j'_{n'}(kr_M)}{h'_{n'}(kr_M)}h_{n'}(kr_M) \right)
\end{eqnarray}\label{eq:shy2}
are coefficients of the radial functions that correspond to the SLA and SMA array types, respectively \cite{rafaely2004plane,rafaely2011optimal}, where
$\rho$ is the air density, 
$c$ is the speed of sound, 
$i^2 = -1$, 
$j_{n}(\cdot)$ and $j'_{n}(\cdot)$ are the spherical Bessel function of order $n$ and its derivative, respectively,
and $h_{n}(\cdot)$ and $h'_{n}(\cdot)$ are the spherical Hankel function of the first kind of order $n$ and its derivative, respectively. 
In particular, $b_{n}^L(kr_L)$ and $b_{n'}^M(kr_M)$ from Eqs.~(\ref{eq:shy1}) and (\ref{eq:shy2}) are defined in accordance with the steady state solution from \cite{williams1999fourier}, employing a Fourier basis with negative frequency \cite{tourbabin2015consistent}.
$Y_{n}^{{m}}(\bm \beta_0)$ is the SH function of order $n$ and degree $m$, evaluated at elevation angle $\omega_{0}$ and azimuth angle $\psi_{0 }$, abbreviated as $\bm \beta_0 = (\omega_{ 0 }, \psi_{0})$, which point at the SMA position with respect to the SLA center. 
$Y_{n}^{{m}}(\bm \beta_0)$ is given by \cite{williams1999fourier}: 
\begin{eqnarray}\label{eq:shy3}
Y_{n}^{{m}}(\bm \beta_0) & = &  \sqrt{ \frac{2n + 1}{ 4\pi}  \frac{(n-m)!}{(n + m)!}  } P_{n}^{m}(\cos\omega_{0}) e^{im \psi_{0 }},\end{eqnarray}\label{eq:shy2}
where $(\cdot)!$ denotes the factorial function and $P_{n}^{m}(\cdot)$ are the associated Legendre functions.  
Similarly for the SMA, the SH function $Y_{n'}^{{m'}}(\bm \xi_0)$ is evaluated at elevation and azimuth angles $\bm \xi_0= (\theta_{0 }, \phi_{0 })$, which point at the SLA position with respect to the SMA center. 
$\bm \beta_{0 }$ and $\bm \xi_0$ are referred to as the direction of radiation (DOR) and direction of arrival (DOA), respectively. 
Finally, $r_0$ is the distance between the array centers.  
Note that for free-field, $\bm H(k)$ has inherently unit rank for all $k$.
However, in a reverberant room $\bm H(k)$ is expected to be of full rank. 
For a discussion on the properties of $\bm H(k)$ the reader is referred to \cite{morgenstern2015theory}. 
}

\begin{figure}
\centering
\includegraphics[trim = 0mm 0mm 0mm 5mm, clip, width=22pc]{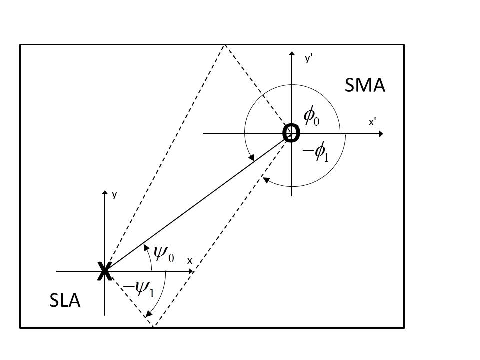}
\caption{System diagram in the x-y plane; X and O represent the SLA and SMA, respectively. The solid line represents direct sound and the dashed lines represent reflections (only two reflections are illustrated). $ \psi_{0 }$ and $ \phi_{0 }$ are the azimuth angles of the DOR and DOA for the direct sound, respectively. $ \psi_{1 }$ and $ \phi_{1 }$ are the corresponding angles for the sound reflected by the wall at the bottom of the figure. In this case, $\omega_{0 } = \theta_{0 } = \omega_{1 } = \theta_{1 } =90^\circ$.}
\label{fig0}
\end{figure}

For a system positioned in a room, a MIMO RTF matrix, $\bm H(k)$, is presented as a summation of MIMO RTF matrices for different room reflections \cite{allen1979image}, given by: 
\begin{eqnarray}\label{eq:base4}
\bm H(k) & =& \sum_{g} a_g(k) \bm h^L_g(k) [\bm h^M_g(k)]^\mathrm{H}, 
\end{eqnarray} 
where $\bm h^L_g(k)$ and $\bm h^M_g(k)$ are the free-field SLA and SMA vectors for reflection $g$, respectively, and $a_g(k)$ accounts for that reflection's attenuation due to absorption by the walls.  
$\bm h^L_g(k)$ is as in Eq.~(\ref{eq:bbase01}), but using the corresponding acoustic path length, $r_g$, instead of $r_0$. 
\textcolor{hai}{
It is important to note that when employing the image source method with a directional SLA, the image sources do not only displace but they also mirror according to the corresponding reflection path.   
To account for this, $\bm \beta_g$ is substituted in Eq.~(\ref{eq:bbase01}), instead of $\bm \beta_0$, denoting the DOR of the $g^{th}$ mirrored and displaced image source. 
The exact mirroring is determined by the respective reflection path. 
Similarly, $\bm h^M_g(k)$ is as in Eq.~(\ref{eq:bbase02}), but using $\bm \xi_g$, the DOA of the $g^{th}$ image source. In this case, $\bm \xi_g$ is deteremined only by the displacement of the image source. 
It is also important to note that when positioned in a room, the rank of $\bm H(k)$ is bounded by the number of significant reflections or by the dimensions of $\bm H(k)$ (the lower of the two); i.e., $min(I, (N_L+1)^2, (N_M+1)^2)$, where $I$ is the number of significant reflections. 
For further details the reader is referred to \cite{morgenstern2015theory}. 
}
Figure \ref{fig0} shows a schematic illustration of the system in a room. 
For simplicity, a system diagram is given in the x-y plane for which both of the arrays are positioned at the same height. 
The coordinate systems of the arrays are aligned with the walls of the room and, therefore, the DOA and DOR elevation angles are both set at $\omega_{0 } = \theta_{0 } = 90^\circ$. 
In the figure, the solid line represents the direct sound, and $ \psi_{0 }$ and $ \phi_{0 }$ are the corresponding azimuth angles of the DOA and DOR, respectively. 
The dashed lines represent two wall reflections, and  $ \psi_{1}$ and $ \phi_{1}$ are the corresponding azimuth angles of one of these reflections.  

The system is now reformulated with normalization applied to the arrays. 
\textcolor{hai}{In \cite{rafaely2011optimal} a normalization method has been proposed for spherical arrays, where the plane-wave decomposition (PWD) representation of the sound field is calculated given the pressure on a sphere. 
Applying the PWD to impulse response information effectively removes the acoustic effects of an array's physical construction, and facilitates accurate DOA estimations \cite{park2005sound}. 
In \cite{rafaely2011optimal} a beamforming framework was proposed for SLAs, in which an equivalent to the PWD is applied for obtaining the far-field radiation pattern of an SLA.
In particular, in \cite{morgenstern2015theory} it was shown that normalization leads to improved, frequency-independent performance in a room acoustics application.}
A normalized MIMO RTF matrix, $\bm G(k)$, is defined, given $\bm H(k)$, as: 
\begin{eqnarray}\label{eq:model7}
\bm H(k) & =& \bm B_L(k) \bm G(k)  \bm B_M(k),
\end{eqnarray} 
where 
\begin{eqnarray}\label{eq:model8}
\bm B_L(k) & = &  \text{diag} [ b_{0}^L(kr_L),\, b_{1}^L(kr_L), \,b_{1}^L (kr_L), \,b_{1}^L(kr_L),  ... , \,b_{N_L}^L(kr_L) ] \,\, \text{and} \\
\bm B_M(k) & = & \text{diag} [ b_{0}^M(kr_M), \,b_{1}^M(kr_M),\,b_{1}^M(kr_M),\,b_{1}^M(kr_M),  ... , \,b_{N_M}^M(kr_M) ]\end{eqnarray}\label{eq:model9}
have dimensions of $(N_L+1)^2 \times (N_L+1)^2$ and $(N_M+1)^2 \times (N_M+1)^2$, respectively. 
\textcolor{hai}{The system output is now formulated as: 
\begin{eqnarray}\label{eq:tired}
y(k) & = & \bm \gamma(k)^{\mathrm{H}}   { \bm G}(k) {\bm \lambda}   s(k),
\end{eqnarray}
where ${\bm \gamma}(k) =[\bm B_L(k)]^{\mathrm{H}} \bar{\bm \gamma}(k) $ and ${\bm \lambda}(k) =\bm B_M(k) \bar{\bm \lambda}(k)$. 
}
Finally, note that since $\bm B_L(k)$ and $\bm B_M(k)$ introduce ill conditioning at low frequencies, it is recommended to employ robust methods for inverting this matrix \cite{li2007flexible}. 

Three SMA beamforming vectors are used in this paper. 
First, an omnidirectional directivity is set at the SMA, which leads to the system output representing the sound pressure at the center of the SMA. 
This is required for the derivation of SLA beamforming methods in the next section. 
For this case, the system output is formulated as in Eq.~\eqref{eq:tired}, using beamforming vector $\bm \lambda^{O}$, defined as: 
\begin{eqnarray}\label{eq:model6}
\bm \lambda^{O}& =&  [1,\,0,\,0,...,0]^{\mathrm T}. 
\end{eqnarray} 
Second, the PWD of the sound field surrounding the SMA is calculated, and is shown in Sec.~\ref{sec:simulationstudies} for spatial angles $\bm \xi_q$, with respect to the SMA center. 
This is done using beamforming vector \cite{rafaely2004plane}:   
\begin{eqnarray}\label{eq:model6a}
\bm \lambda^{P}& =&  \begin{bmatrix}Y_{0}^{0}(\bm \xi_q), Y_{1}^{-1}(\bm \xi_q), Y_{1}^{0}(\bm \xi_q), ..., Y_{N_M}^{N_M}(\bm \xi_q) \end{bmatrix}^{\mathrm H}. 
\end{eqnarray} 
Third, binaural responses are synthesized for listening tests in Sec.~\ref{sec:listeningtests}, using vector $\bm \lambda^{B}$, defined as: 
\begin{eqnarray}\label{eq:model6b}
\bm \lambda^{B}& =& \begin{bmatrix}Q_{00}^{l}(k),\,Q_{1(-1)}^l(k),\,Q_{10}^x(k),...,Q_{N_MN_M}^l(k)\end{bmatrix}^{\mathrm H},  
\end{eqnarray}
where $Q_{nm}^{l}(k)$ are coefficients that account for the head-related transfer functions (HRTFs) of the left ear\cite{rafaely2010interaural}. 
For syntehsizing the response for the right ear, $\bm \lambda^{B}$ will be applied using $Q_{nm}^{r}(k)$, which are the HRTFs of the right ear. 

Finally, room-impulse-response (RIR) matrices are formulated for the system. 
The RTF system matrix $\bm G(k)$ is sampled in the frequency domain, and the RIR is computed by employing the inverse discrete Fourier transform (DFT).
For convenience, RIR matrices are written using the same notation as that of the RTF matrices, but with the dependance on the continuous wavenumber $k$ changed to a dependance on the discrete-time index $t$. 
\textcolor{hai}{At this stage, the normalized beamforming coefficients are assumed to be constant in frequency, so as to facilitate a simple formulation of the system output. 
Therefore, the dependancy of $\bm \gamma$ and $\bm \lambda$ on $k$ is henceforth omitted. 
}
Under this assumption, the RIR, $h[t]$, is formulated for the system as in Eq.~(\ref{eq:tired}), but using $\bm G[t]$, instead of $\bm G(k)$, as: 
\begin{eqnarray}\label{eq:tired2}
h[t] & = & {\bm \gamma}^{\mathrm{H}}   { \bm G}[t] {\bm \lambda}. 
\end{eqnarray} 
\textcolor{hai}{Note that $\bm \lambda^{B}$ from Eq.~(\ref{eq:model6b}) is frequency variant by definition. 
In Sec.~\ref{sec:listeningtests}, the application of beamforming with $\bm \lambda^{B}$ after computing frequency-independent SLA beamforming coefficients will be discussed in detail.} 

\section{SPATIAL MODIFICATION OF REVERBERATION}\label{sec:methods}

In this section, methods for spatial dereverberation and reverberation are proposed.
These methods involve the optimization of room-acoustics measures with respect to the SLA beamforming coefficients vector.
\textcolor{hai}{
The formulation introduced in the previous section, which assumed frequency invariant beamforming coefficients, leads to simple solutions, which also force the system to focus on spatial modifications of the sound field.  
The extension of the proposed methods to frequency dependent beamformers is proposed for future work. 
}

\subsection{Direct-to-reverberant ratio}\label{sec:DRR}

A widely-used objective measure for evaluating the level of reverberation of an RIR is the direct-to-reverberant ratio (DRR) \cite{kinsler1999fundamentals,naylor2010speech}. 
\textcolor{hai}{The DRR, denoted as $DRR$, is defined as the power of the direct sound relative to the power of the reverberant sound within an RIR, $h[t]$, and is given by:} 
\begin{eqnarray}\label{eq:method1}
DRR & =&  10\, \text{log}_{10 }\frac{\sum_{t=0}^{t_D} |h[t]|^2 }{\sum_{t= t_D}^{\infty} |h[t]|^2 },  
\end{eqnarray} 
where $t_D$ is chosen such that the time range $0\leq t\leq t_D$ refers to the segment of the direct sound in the RIR, and the time range $t>t_D$ refers to that of the reflected sound. 
For a MIMO system, the DRR can be formulated by applying beamforming to the arrays; 
this is implemented by substituting $h[t]$ from Eq.~(\ref{eq:tired2}) in Eq.~(\ref{eq:method1}). 
This is denoted $DRR^{MIMO}$ and is written: 
\begin{eqnarray}\label{eq:method2}
DRR^{MIMO} & = &   
\frac{ \bm\gamma^{\mathrm{H}} \bm A \bm\gamma }
{   \bm\gamma^{\mathrm{H}} \bm B \bm\gamma }, 
\end{eqnarray} 
where 
\begin{eqnarray}\label{eq:method25}
\bm A & = &  
 \sum_{t=0}^{t_D}  
\bm G[t]  \bm \lambda^{O}
\begin{bmatrix} \bm \lambda^{O}\end{bmatrix}^\mathrm{H} [\bm G[t] ]^\mathrm{H}
, \\
\bm B & = &  
\sum_{t=t_D}^{\infty}  
\bm G[t]  \bm \lambda^{O}
\begin{bmatrix} \bm \lambda^{O}\end{bmatrix}^\mathrm{H} [\bm G[t]]^\mathrm{H}.
\label{eq:method26}
\end{eqnarray}
In particular, $\bm \lambda^{O}$ is used for the SMA and a general beamforming vector $\bm \gamma$ is set for the SLA. 
\textcolor{hai}{Also, in Eq.~(\ref{eq:method25}) it is assumed that $\bm G[t]$ is causal. 
Firstly, $h[t]$ from Eq.~(\ref{eq:tired2}) represents the response to the input of the SLA at the output of the SMA, which describes a causal system, so that $h[t]$ is causal. 
The functions composing $h[t]$, i.e., $\bm G[t]$, are not proven here to be causal. 
In practice, these were, in fact, found to be causal, probably due to the long delay imposed by the distance between the SLA and the SMA. 
$\bm \gamma$ that maximizes and minimizes $DRR^{MIMO}$, referred to as $maxDRR$ and $minDRR$, respectively, can be computed by solving
\begin{eqnarray}\label{eq:endend1}
maxDRR & = & \underset{\bm\gamma}{\text{arg max}}\,\, DRR^{MIMO} \,\,\,\, \text{and} \\
minDRR & = & \underset{\bm\gamma}{\text{arg min}} \,\,DRR^{MIMO},\label{eq:endend2}
\end{eqnarray}
respectively. 
In particular, $DRR^{MIMO}$ in Eq.~(\ref{eq:method2}) can be formulated as a generalized Rayleigh quotient.
Following the discussion in \cite{morgenstern2015theory}, since $\bm B$ is constructed as a summation of $\bm G[t]$ for multiple reflections, it can be assumed to be invertible for a reverberant room with many significant reflections. 
Therefore, $maxDRR$ and $minDRR$ can be computed as the eigenvectors that correspond to the maximum and minimum generalized eigenvalues of $\bm A \bm \gamma =\upsilon \bm B \bm \gamma $, respectively, where $\bm A$ and $\bm B$ are the matrices defined in Eqs.~(\ref{eq:method25}) and (\ref{eq:method26}), respectively, and $\upsilon$ is the eigenvalue. 
Reverberation characteristics can now be modified by applying these beamformers, as will be shown in Sec.~\ref{sec:simulationstudies}. }

\subsection{Early-to-late energy index}\label{sec:clarity}

The DRR is known to affect speech intelligibility. 
This is because the direct sound is considered as an intelligible signal, and reverberation as noise. 
In the human auditory system, however, early reflections also contribute to the intelligibility of a speech signal; 
early reflections are perceived as a coloration of the direct sound component if they arrive not later than about 50 ms after the direct sound \cite{naylor2010speech}.
An objective measure for evaluating the energy of these early reflections, compared to the reverberation tail, is the early-to-late index, also referred to as clarity or $C50$ \cite{standard19973382}. 
$C50$ is defined using Eq.~(\ref{eq:method1}), by replacing $t_D$ from the limits of the summations (in both numerator and denominator), with $t_C$. 
The latter is chosen such that $0\leq t\leq t_C$ refers to the segment in the RIR of the direct sound and reflections that arrive no later than $50\,$ms after it. 
For a MIMO system, $C50^{MIMO}$ can be formulated as in Eq.~(\ref{eq:method2}), but using $\bm A$ and $\bm B$ that are defined using $t_C$, instead of $t_D$, in Eqs.~(\ref{eq:method25}) and (\ref{eq:method26}), respectively. 
\textcolor{hai}{
$\bm \gamma$ that minimizes and maximizes $C50^{MIMO}$, referred to as $maxC50$ and $minC50$, respectively, can be computed by solving the optimization problems in Eqs.~(\ref{eq:endend1}) and (\ref{eq:endend2}), but using $C50^{MIMO}$ instead of $DRR^{MIMO}$. 
Again, since $C50^{MIMO}$ can be formulated as a generalized Rayleigh quotient, $maxC50$ and $minC50$ can be computed as the eigenvectors that correspond to the maximum and minimum generalized eigenvalues of $\bm A \bm \gamma = \upsilon \bm B \bm \gamma$, respectively.    
Here, suitable $\bm A$ and $\bm B$ that have been defined using $t_C$ are used. 
As in the case of $DRR^{MIMO}$, reverberation characteristics can be modified by applying these beamformers, as will be shown in Sec.~\ref{sec:simulationstudies}.}  

\textcolor{hai}{Optimization with respect to $DRR$ and $C50$ was presented in this section, since these are used for quantifying reverberation. 
In practice, optimization of $DRR$ and $C50$ may lead to coloration and an unnatural decay pattern of the output signal. 
Further studies on the optimization formulation that takes these effect into account and that is more appropriate in the context of listener perception e.g.,~\cite{mertins2010room,jungmann2012combined,lim2014robust}, is proposed for further study. }

\section{Simulation studies}\label{sec:simulationstudies}

In this section, simulations are presented for sound fields in a room, produced by an SLA that employs the proposed methods. 
First, the effect of these methods on the \textcolor{hai}{spectro-temporal} attributes of reverberation are studied for various room and SLA parameters through the analysis of RIR energy decay curves (EDCs) recorded by an omnidirectional microphone.   
Then, $C50$ is evaluated in a similar manner to study the effects on clarity. 
In addition, $T_{20}$, a measure for evaluating the reverberation time (RT) of a response \cite{standard19973382}, is calculated with the aim of studying the effects on reverberation. 
The employment of the proposed beamformers is expected to change spatial characteristics of the sound fields. 
To study these effects, the plane-wave amplitude distributions of the sound field that surrounds the omnidirectional microphone are studied. 
This is done by replacing that microphone with an SMA, which, together with the SLA, comprises a MIMO system. 
\textcolor{hai}{Finally, since the beamformers are designed assuming perfect knowledge of the RIR matrix, an analysis of robustness is presented for cases where such knowledge is not completely available.} 

\subsection{Setup}\label{sec:simulationsetup}

An analysis is provided for two system configurations and for a large- and a medium-sized rooms, with two different RTs.  
For both system configurations, the same pair of SLA and SMA are simulated, with radii of $r_L = 0.20\,$m and $r_M = 0.12\,$m, respectively.
In particular, several values of the SLA SHs orders are used, $N_L = 2,3,$ and 4, while the SMA SHs order is set to $N_M= 5$. 
\textcolor{hai}{These orders are chosen to roughly reflect the orders of currently available SLA and SMA systems \cite{avizienis2006compact,meyer2002highly}.}
For the first system configuration, the SLA and SMA are positioned at Cartesian coordinates $(5,8,3)\,$m and $(22,19,7)\,$m, respectively, within a large room, referred to as $room\,1$, with dimensions of $(44,24,13)\,$m. 
Two different RTs of $1.14\,$s and $0.71\,$s are simulated for this room, with absorption coefficients  for all walls and frequencies set to $0.5$ and $0.8$, respectively. 
For the second system configuration, the SLA and SMA are positioned at Cartesian coordinates $(14,8,5)\,$m and $(5,3,2)\,$m, respectively, within a smaller room, referred to as $room\,2$, with dimensions of $(30,13,7)\,$m. 
For this room, two different RTs of $0.80\,$s and $0.45\,$s are simulated using absorption coefficients for all walls and frequencies of $0.4$ and $0.7$, respectively. 
Multiple transfer functions were simulated using the McRoomSim software \cite{wabnitz2010room} for each element of the MIMO RTF matrix given in Eq.~(\ref{eq:base4}), using a sampling frequency of $48\,$kHz. 
Before transforming these functions to RIRs, they are band-pass filtered using an upper frequency of $5.66\,$kHz and a lower frequency of $300\,$Hz. 
Filtering is applied since in practice, real systems are expected to include significant errors outside of this frequency range due to spatial aliasing and model mismatch \cite{morgenstern2015joint}. 

\subsection{Methods}\label{sec:simulationmethods}

\textcolor{hai}{Beamforming coefficients were designed for the systems configurations and the SLA SHs orders.  
$maxDRR$ and $minDRR$ were designed as detailed in Sec.~\ref{sec:DRR}, and $maxC50$ and $minC50$, as detailed in Sec.~\ref{sec:clarity}. 
Additional beamforming vectors are used in this section for comparison.
These include: a) an omnidirectional source, modelled by applying $\bm \gamma = [1, 0, ...0 ]^{\mathrm{T}}$ to the SLA in Eq.~(\ref{eq:bcase0}) and referred to as $omni$ in this section;  
b) a maximum directivity beamformer \cite{van2004detection}, with its look direction set at the SMA direction.
This beamformer will be used for several SLA SHs orders throughout the section, and is referred to as $maxFIX$; 
c) an $N^{th}$-order beamformer with a Cardioid beam pattern \cite{elko2004differential}, extended for several SHs orders. The look direction of this beamformer is set to be opposite to the SMA position, such that its null points in the SMA direction. This beamformer is referred to as $minFIX$.}  

\subsection{Results}\label{sec:simulationresults}
 
\begin{figure}
\centering
\includegraphics[width=23pc]{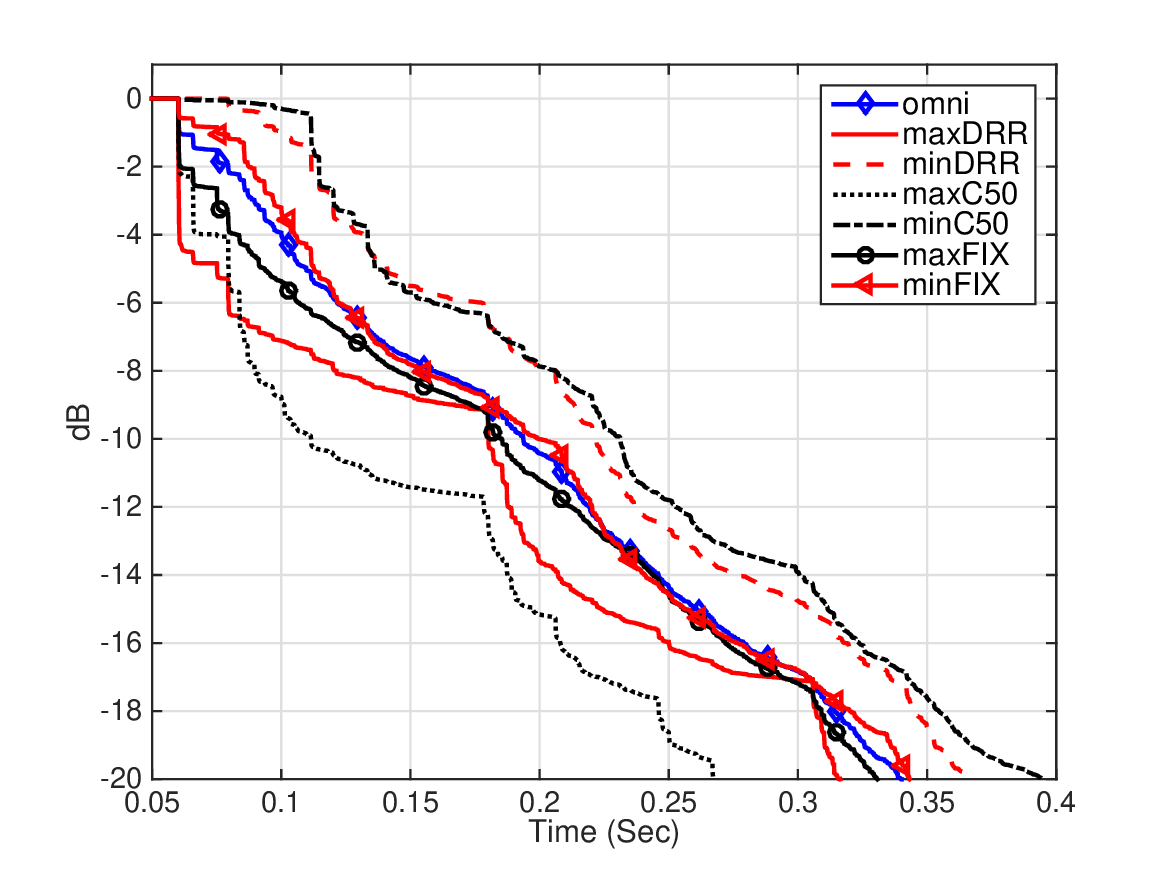}
\caption{EDCs for $room\,1$ with RT $=1.14\,$s, for SLA beamforming vectors $maxDRR$, $minDRR$, $maxC50$, $minC50$, $maxFIX$, and $minFIX$, employing SLA SHs order $N_L = 2$, and beamforming vector $omni$.}
\label{fig:H_r2_0.4}
\end{figure}

The EDC of an RIR represents the decay in the sound pressure level after a sound source has been stopped \cite{standard19973382}. 
EDCs are calculated for $room\,1$ using the Schroeder integration method \cite{schroeder1965new} with $\bm \lambda^O$ used at the SMA and for $maxDRR$, $minDRR$, $maxC50$, and $minC50$, employing an SLA SHs order $N_L = 2$.   
These are presented in Fig.~\ref{fig:H_r2_0.4}. 
\textcolor{hai}{The EDCs that correspond to $maxFIX$ and $minFIX$ with $N_L = 2$ and to $omni$ are also shown for reference.} 
It is evident that all curves have similar slopes from $200\,$ms onwards. 
In particular, the differences between these curves are mainly seen at times that correspond to the early reflections.  
$maxDRR$ shows a drop of almost $5\,$dB just after the time delay of the direct sound (about $60\,$ms). 
$50\,$ms after that time delay, at $110\,$ms, which is also the value used for $t_c$, the level of $maxC50$ is the lowest between the curves.
On the other hand, $minDRR$ and $minC50$ maintain higher amplitude in the decay curves, with the level for $minC50$ evaluated at about $-2\,dB$ at $t_C$. 
This implies that less energy is found in the first 50$\,$ms of the response, compared to the other curves. 
\textcolor{hai}{The behavior of both $maxFIX$ and $minFIX$ is similar to that of $omni$, with minor differences in dB just after the time delay of the direct sound.} 

\begin{figure}
\centering
\includegraphics[width=23pc]{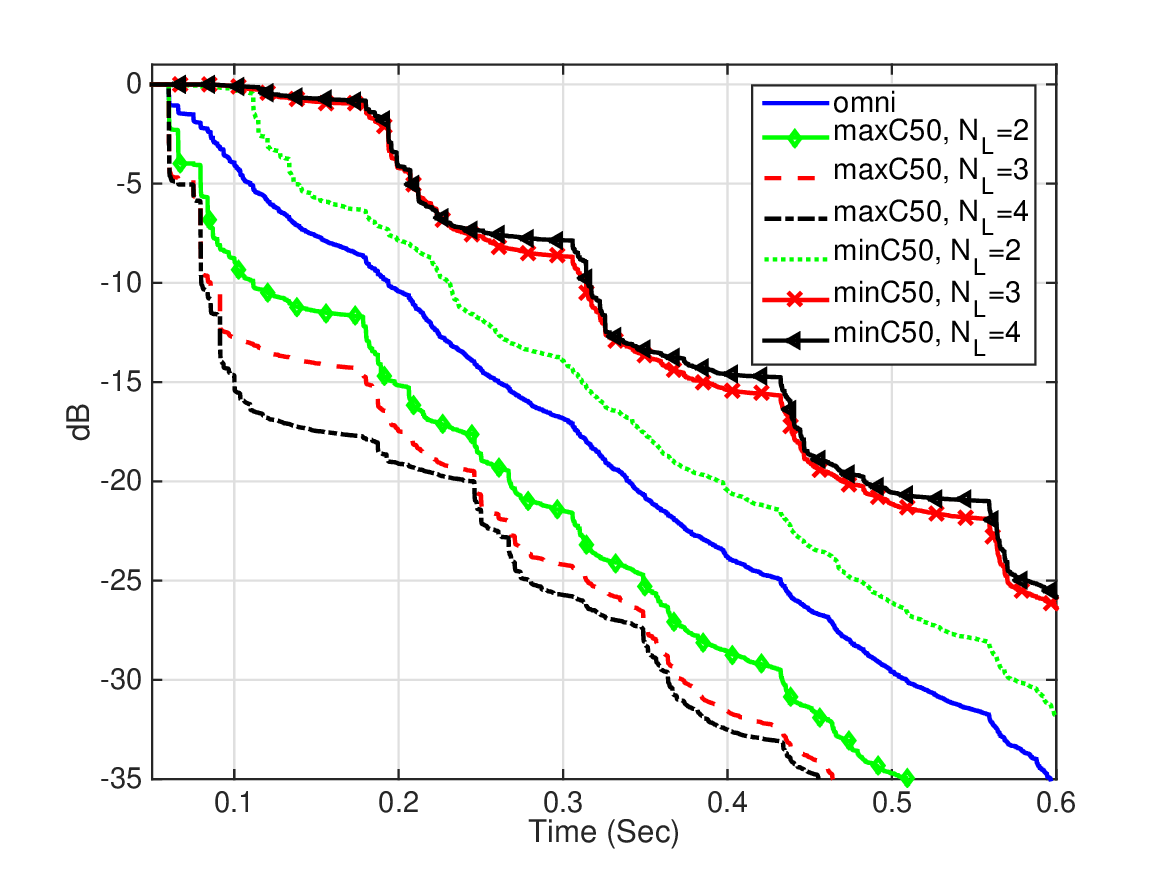}
\caption{EDCs for $room\,1$ with RT $=1.14\,$s, for SLA beamforming vectors $maxC50$ and $minC50$, employing SLA SHs orders $N_L = 2,3,$ and $4$, and beamforming vector $omni$.}
\label{fig:H_r2_0.4_NLvariation}
\end{figure}

To study the effects of the spatial resolution of the SLA on these decay curves, Fig.~\ref{fig:H_r2_0.4_NLvariation} shows the EDCs for $maxC50$ and $minC50$ for several SLA SHs orders, as in Fig.~\ref{fig:H_r2_0.4}. 
The EDC for $omni$ is also presented for reference. 
Employing a higher SHs order at the SLA results in modifications for both $maxC50$ and $minC50$. 
For $maxC50$, higher attenuation of the decay curves at $t_C$ is evident for higher orders. 
For $minC50$, all decay curves maintain a high amplitude until $t_C$. 
The similar behavior of curves corresponding to $N_L = 3$ and $4$, implies that the additional spatial resolution is not beneficial for the the distribution of the reflections' time dealys, and the DORs and DOAs at the SLA and SMA, respectively, in the room.  

The variation in clarity due to the employment of the SLA beamforming vectors is studied for a wide range of conditions. 
$C50$ values are calculated using the EDCs for both rooms, with two different RTs, and for selected SLA beamforming vectors and several SLA SHs orders.   
These are presented in Table \ref{table:clarity2}. 
\textcolor{hai}{
\begin{table}
\centering
\small
\begin{tabular}{ | l  l | c | c | c | c | }
\hline
\multicolumn{2}{ | c | }{Beamformer and}   & $room\,1$ & $room\,1$ & $room\,2$ & $room\,2$\\
\multicolumn{2}{  | c | }{SLA order} & ($1.14\,$s) & ($0.71\,$s) & ($0.80\,$s) & ($0.45\,$s) \\
\hline 
\multicolumn{2}{  | l | }{ $omni$}  & 2.98 & 11.32 & 3.44 & 12.60\\
\hline
$maxC50$, &$N_L = 2$ & 9.78 & 18.50 & 14.07 & 24.50 \\
 &$N_L = 3$ &  12.92 & 23.05 & 17.43 & 30.03 \\
&$N_L = 4$ &  16.38 & 27.39 & 20.35 & 33.24\\
\hline
$minC50$, &$N_L = 2$ & -4.02 & 1.12 & -10.82 & -8.73\\
 &$N_L = 3$ & -12.40 & -9.22 & -12.95 & -11.37 \\
 &$N_L = 4$ & -15.12 & -12.43 & -15.09 & -13.64\\
\hline
$maxFIX$, &$N_L = 2$ & 4.48 & 13.54 & 10.31 & 19.70 \\
 &$N_L = 3$ &  5.82 & 14.91 & 12.15 & 21.89\\
&$N_L = 4$ & 7.14 & 16.44 & 14.02 & 24.26\\
\hline
$minFIX$, &$N_L = 2$ & 2.23 & 9.72 & 0.97 & 8.53\\
 &$N_L = 3$ & 1.96 & 9.34 & 0.57 & 7.98 \\
 &$N_L = 4$ & 1.69 & 9.02 & 0.35 & 7.68\\
\hline
\end{tabular}
\caption{$C50\,$[dB] for both rooms (room RT in parentheses), for $maxC50$, $minC50$, $maxFIX$, and $minFIX$ with $N_L = 2,3$ and 4, and for $omni$.}
\label{table:clarity2}
\end{table}
}
$maxC50$ and $minC50$ are seen to significantly increase and decrease, respectively, the value of $C50$, compared to the value of $C50$ with an omnidirectional source.  
For example, when $maxC50$ is applied with $N_L=4$, $C50$ increases by more than 16$\,$dB and 20$\,$dB for $room\,1$ (RT = 0.71$\,$s) and $room\,2$ (RT = 0.45$\,$s), respectively, compared to $omni$. 
Similarly, when $minC50$ is applied using the same SHs order, $C50$ decreases by more than 23$\,$dB and 26$\,$dB for $room\,1$ (RT = 0.71$\,$s) and $room\,2$ (RT = 0.45$\,$s), respectively, compared to $omni$. 
$C50$ can therefore be modified in a range of around $40\,$dB and $46\,$dB for $room\,1$ and  $room\,2$, respectively. 
It is evident that the increase and decrease in $C50$ employing $maxC50$ and $minC50$, respectively, are higher for higher SLA SHs orders, smaller rooms, and higher absorption coefficients. 
This can be explained by the following: 
a) employing a higher SHs order increases the spatial resolution of the array, which enhances its ability to selectively radiate acoustic energy into specific spatial directions of reflections; 
b) typically, the number of reflections in the first 50$\,$ms of an RIR increases for smaller rooms due to the smaller dimensions of the rooms. Therefore, the energy of matrix $\bm A$ of the corresponding generalized eigenvalue problem from Sec.~\ref{sec:clarity} is increased, and higher and lower clarity can be achieved by employing $maxC50$ and $minC50$, respectively; 
c) the energy of these reflections increases relative to the energy of the reverberation tail for higher room absorption coefficients. As in the last case, this leads to higher energy of matrix $\bm A$ of the corresponding generalized eigenvalue problem.
\textcolor{hai}{
$maxFIX$ exhibits similar behavior to $maxC50$, showing an increase in $C50$ for an increase in SHs orders. 
However, the $C50$ values for $maxFIX$ are significantly lower than the values of $maxC50$. 
On the other hand, $minFIX$ shows somewhat different behavior to $minC50$. 
For $minFIX$, higher SHs orders do not necessarly lead to a decrease in $C50$.
Nevertheless, the $C50$ values for $minFIX$ are significantly higher than the values of $minC50$.
}

To study the variations in reverberation due to the employment of the SLA beamforming vectors, $T_{20}$ is calculated using a modified evaluation range of -1$\,$dB to -21$\,$dB on the corresponding EDCs.
$T_{20}$ is chosen for evaluating reverberation since it is based on the early part of the EDC, which is related to a subjective evaluation of reverberation \cite{standard19973382}.
$T_{20}$ values are presented in Table.~\ref{table:T202} for the same rooms, SLA beamforming vectors, and SHs orders as in Table.~\ref{table:clarity2}. 
\begin{table}
\centering
\small
\begin{tabular}{ | l  l | c | c | c | c | }
\hline
\multicolumn{2}{ | c | }{Beamformer and}   & $room\,1$ & $room\,1$ & $room\,2$ & $room\,2$\\
\multicolumn{2}{  | c | }{SLA order} & ($1.14\,$s) & ($0.71\,$s) & ($0.80\,$s) & ($0.45\,$s) \\
\hline 
\multicolumn{2}{  | l | }{ $omni$}  & 0.96 & 0.44 & 0.70 & 0.36\\
\hline
$maxC50$, &$N_L = 2$ &  0.68 & 0.34 & 0.56 & 0.12\\
 &$N_L = 3$ &  0.54 & 0.10 & 0.34 & 0.10\\
&$N_L = 4$ &  0.53 & 0.09 & 0.17 & 0.08\\
\hline
$minC50$, &$N_L = 2$ & 0.90 & 0.61 & 0.65 & 0.37 \\
 &$N_L = 3$ &  0.96 & 0.58 & 0.64 & 0.52\\
 &$N_L = 4$ & 1.14 & 0.42 & 0.64 & 0.51\\
\hline
\end{tabular}
\caption{$T_{20}\,$[s] for both rooms (room RT in parentheses), for $maxC50$ and $minC50$ with $N_L = 2,3$ and 4, and for $omni$.}
\label{table:T202}
\end{table}
It is evident that $maxC50$ yield $T_{20}$ values that are lower than that of $omni$. 
Similarly, $minC50$ yields values that are higher than that of $omni$, with some exceptions (e.g., for $room\,1$ (RT = $1.14\,$s) with $minC50$, employing $N_L = 2$). 
For example, $maxC50$ can lead to a decrease in $T_{20}$ of around $420\,$ms and $520\,$ms for $room\,1$ (RT = $1.14\,$s) and $room\,2$ (RT = $0.80\,$s), respectively, compared to $omni$. 
Similarly, $minC50$ can lead to an increase of around $180\,$ms for some cases, compared to $omni$. 
This behavior can be explained by the inverse relation between RT and clarity; 
beamformers that are optimized to maximize or minimize $C50$ are also expected to lead to a decrease or an increase, respectively, in the RT. 
It is also evident that, in this case, employing a higher SHs order at the SLA does not necessarily increase the modifications in $T_{20}$, as in the case of $C50$.
For example, applying $minC50$ with $N_L = 2$ in $room\,2$ (RT = $0.80\,$s) yields a higher $T_{20}$ compared to $N_L = 4$.   
Possible explanations are that the optimization, applied with respect to $C50$, and not with respect to $T_{20}$, and that the behavior of the EDCs is highly nonlinear. 

\begin{figure}
\centering
\includegraphics[width=23pc]{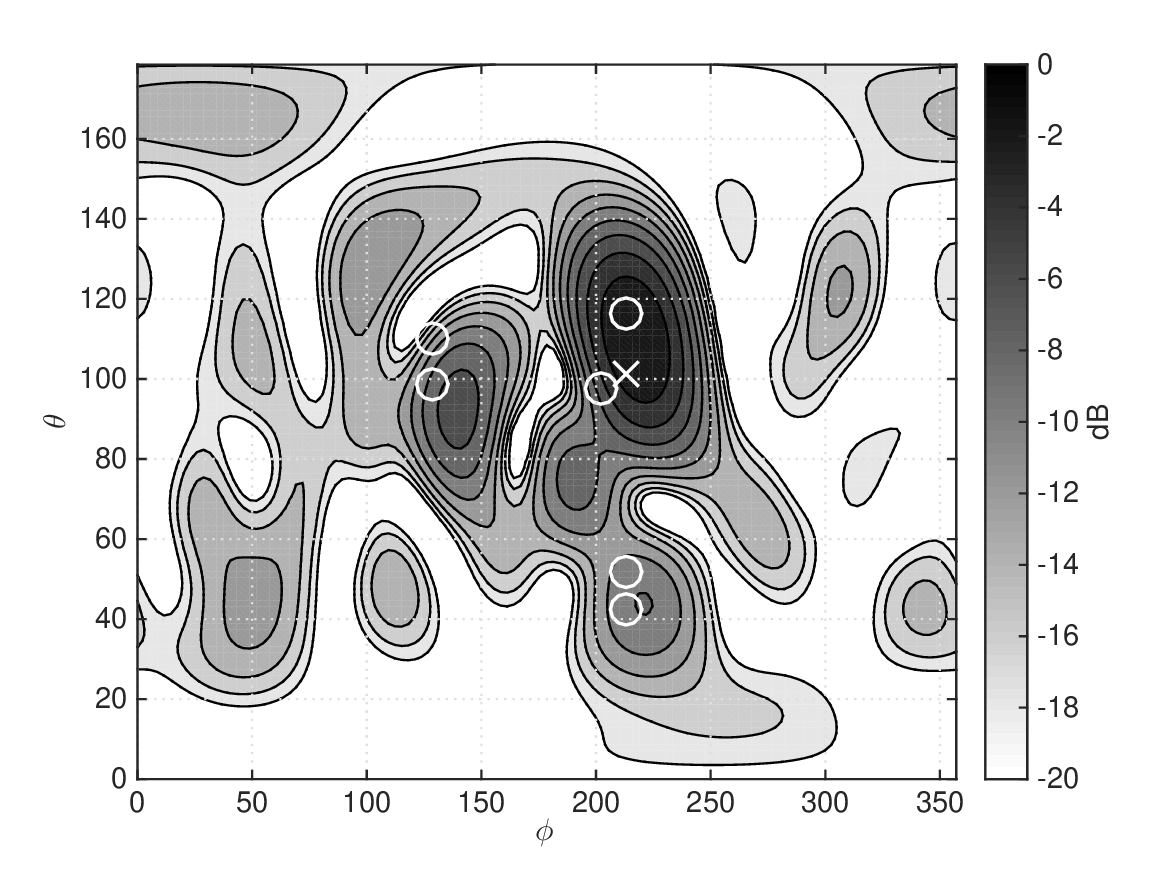}
\caption{PWD around the SMA for $room\,1$, using $N_M = 5$ and for beamforming vector $omni$. The DOAs of direct sound and first order reflections are plotted using `x' and `o's, respectively. } 
\label{fig:SMApwd_omni}
\end{figure}

\begin{figure}
\centering
\includegraphics[width=23pc]{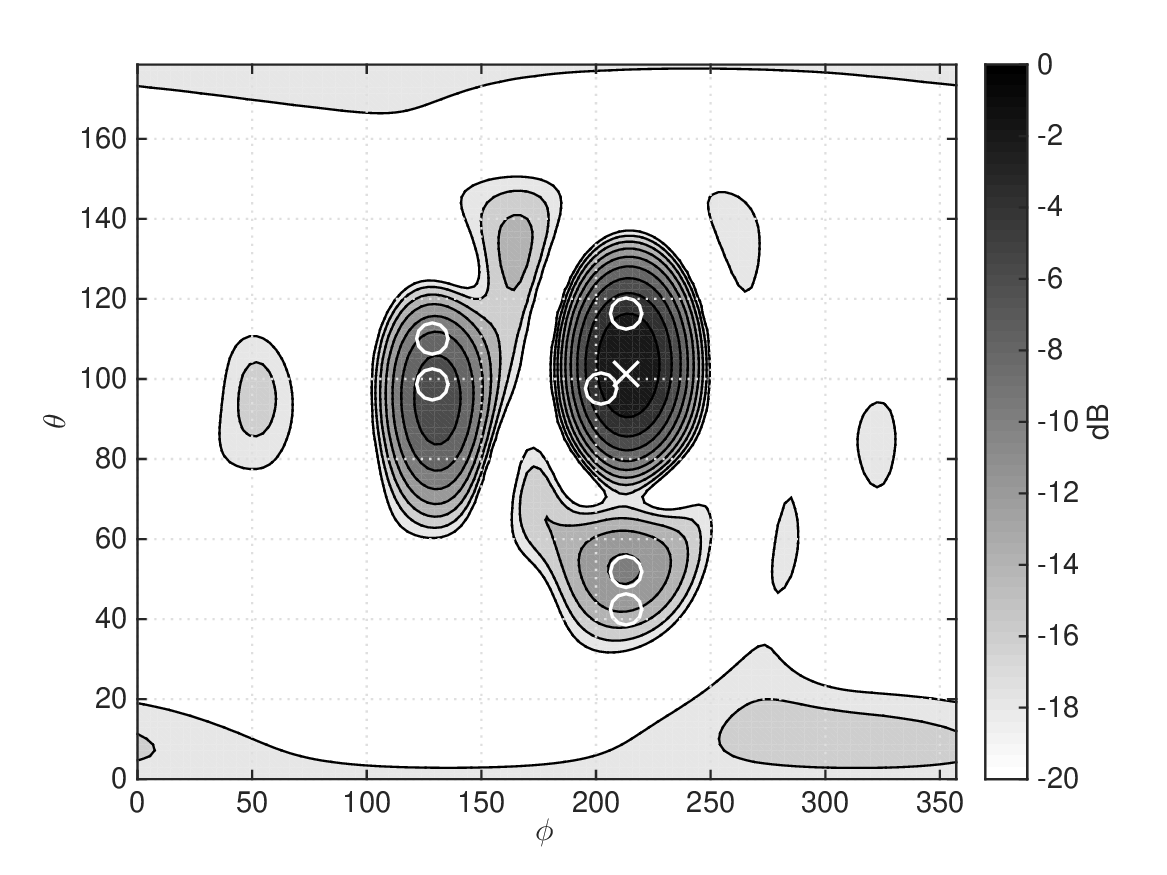}
\caption{Same as Fig.~\ref{fig:SMApwd_omni}, but for $maxC50$ and $N_L = 4$.}
\label{fig:SMApwd_maxDRR}
\end{figure}

\begin{figure}
\centering
\includegraphics[width=23pc]{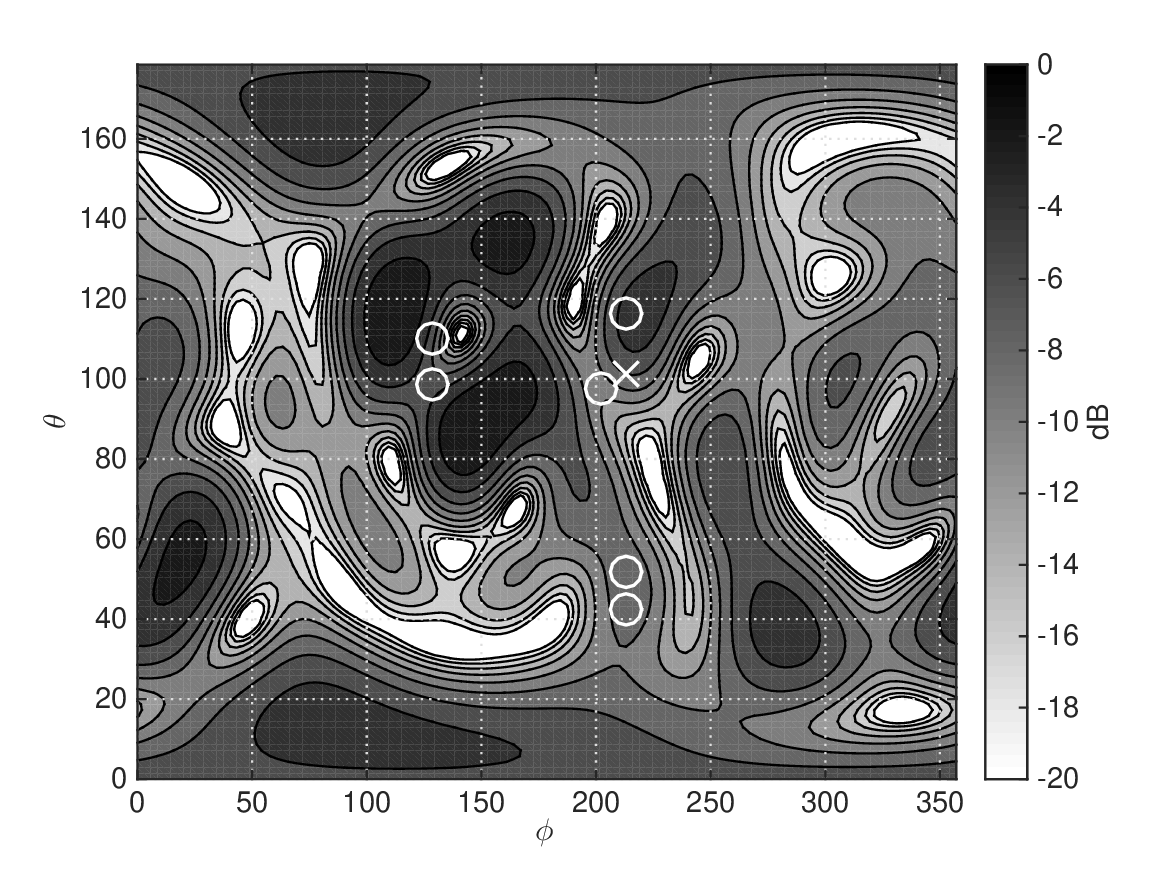}
\caption{Same as Fig.~\ref{fig:SMApwd_omni}, but for $minC50$ and $N_L = 4$.}
\label{fig:SMApwd_minC50}
\end{figure}

Finally, the effects that the beamforming methods have on the spatial characteristics of the sound fields are studied by analyzing of plane-wave amplitude distribution of these sound fields.
The plane-wave amplitude distributions, or directivity functions, are calculated by applying beamforming to both arrays. 
First, $\bm \lambda^{P}$, from Eq.~(\ref{eq:model6a}), is applied to the SMA for a linear distribution of spatial angles, $\bm \xi_q$.
Figure \ref{fig:SMApwd_omni} serves as a reference and presents the PWD around the SMA, for $N_M = 5$ and $f=2.5\,$kHz, due to the employment of an omnidirectional source. 
The PWD is normalized such that the highest magnitude is set at $0\,$dB, and a dynamic range of $20\,$dB is used. 
The DOAs of the direct sound and the first order reflections are also plotted on the figures for reference. 
Figs.~\ref{fig:SMApwd_maxDRR}, and \ref{fig:SMApwd_minC50} present similar PWD plots, but for $maxC50$ and $minC50$, respectively, employing $N_L = 4$.  
It is evident in Fig.~\ref{fig:SMApwd_maxDRR} that for $maxC50$ the sound field directivity shows significant gain, mainly from the DOAs of the early reflections.
On the other hand, in Fig.~\ref{fig:SMApwd_minC50}, the energy of early reflections seems to be attenuated for $minC50$, compared to Fig.~\ref{fig:SMApwd_omni}, while directions away from the direct sound and early reflections seem to be amplified.
This suggests that these methods not only change the \textcolor{hai}{spectro-temporal} sound field attributes, but also perform spatial dereverberation and reverberation. 

In summary, this simulation study has validated the theoretical results, showing that by employing directional sources, instead of an omnidirectional one, a significant modification of the EDCs of RIRs can be achieved. 
Furthermore, employing the proposed beamforming methods has been shown to modify the spatial attributes of the sound field surrounding the SMA. 
It is therefore expected that these differences in the acoustics will also be perceivable by human listeners, as studied in Sec.~\ref{sec:listeningtests}. 

\subsection{Robustness analysis}\label{sec:robustness}

In previous sections, beamformers were developed assuming perfect knowledge of the RIR matrix. 
In practice, however, RIRs cannot be estimated precisely, so that perfect knowledge cannot be available.
Nevertheless, it is important that, in practice, systems are robust to this imperfect knowledge, typically modeled as error in the RIR matrix.
Therefore, an analysis of robustness is presented in this section. 
As an example, $room\,1$ (RT = $1.14\,$s) was chosen for this analysis. 

The characteristics of the actual RIR estimation errors may depend on the system identification method. 
To avoid the need to constrain the analysis to a specific method, a general additive error model is employed in this paper. 
In this model, noise is added directly to the RTF matrix. 
Moreover, since the variance of the actual errors may be frequency-dependant \cite{rafaely2005analysis}, error analysis is performed in octave bands. 
Beamformers are designed based on a given RTF matrix, and then applied to the same RTF matrix, but with the addition of noise. 
The noise added to each element of the RIR was was assumed to be i.i.d. zero-mean white noise. 
The noise variance was chosen to produce 30$\,$dB SNR in octave bands 0.5, 1, 2, and $4\,$kHz. 

\begin{figure}
\centering
   \begin{subfigure}[b]{0.49\textwidth}
   \includegraphics[width=0.93\textwidth]{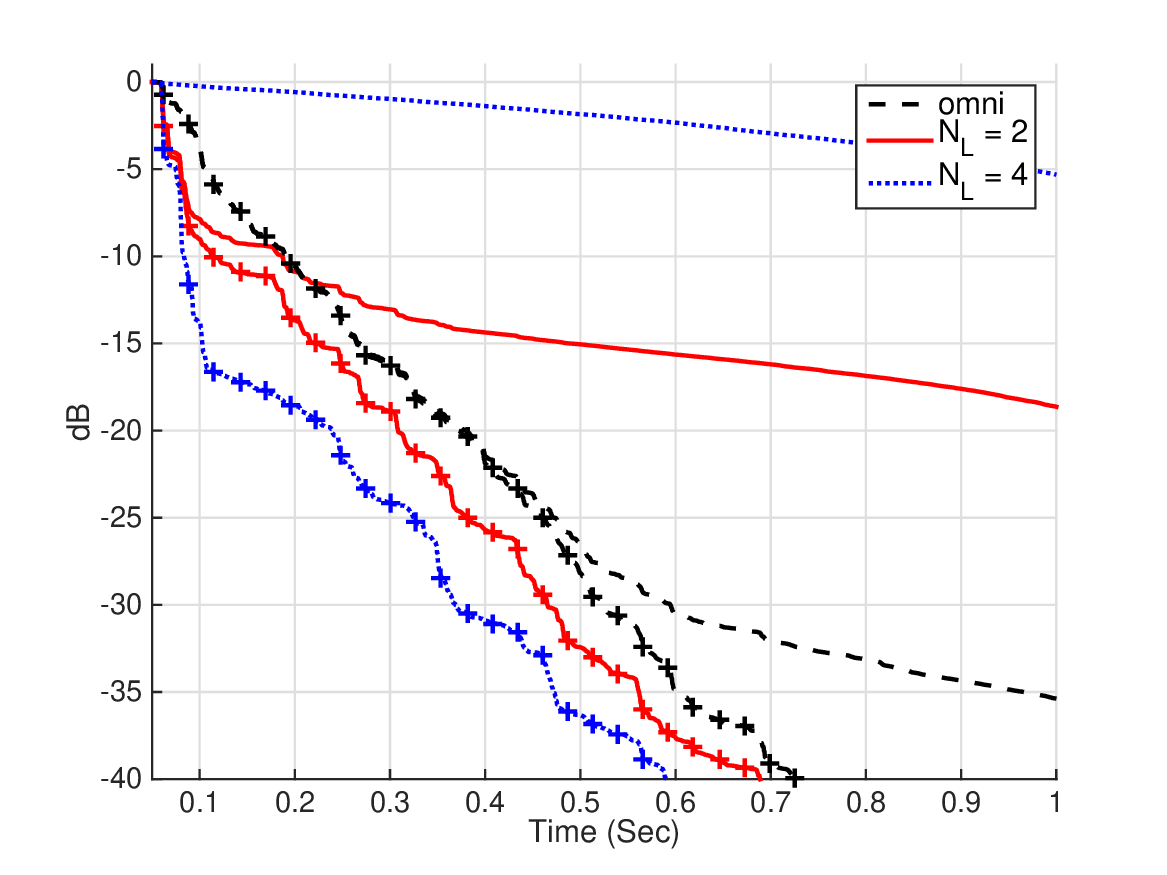}
   \caption{$0.5\,$kHz}
   \label{fig:Ng1} 
\end{subfigure}
\begin{subfigure}[b]{0.49\textwidth}
   \includegraphics[width=0.93\textwidth]{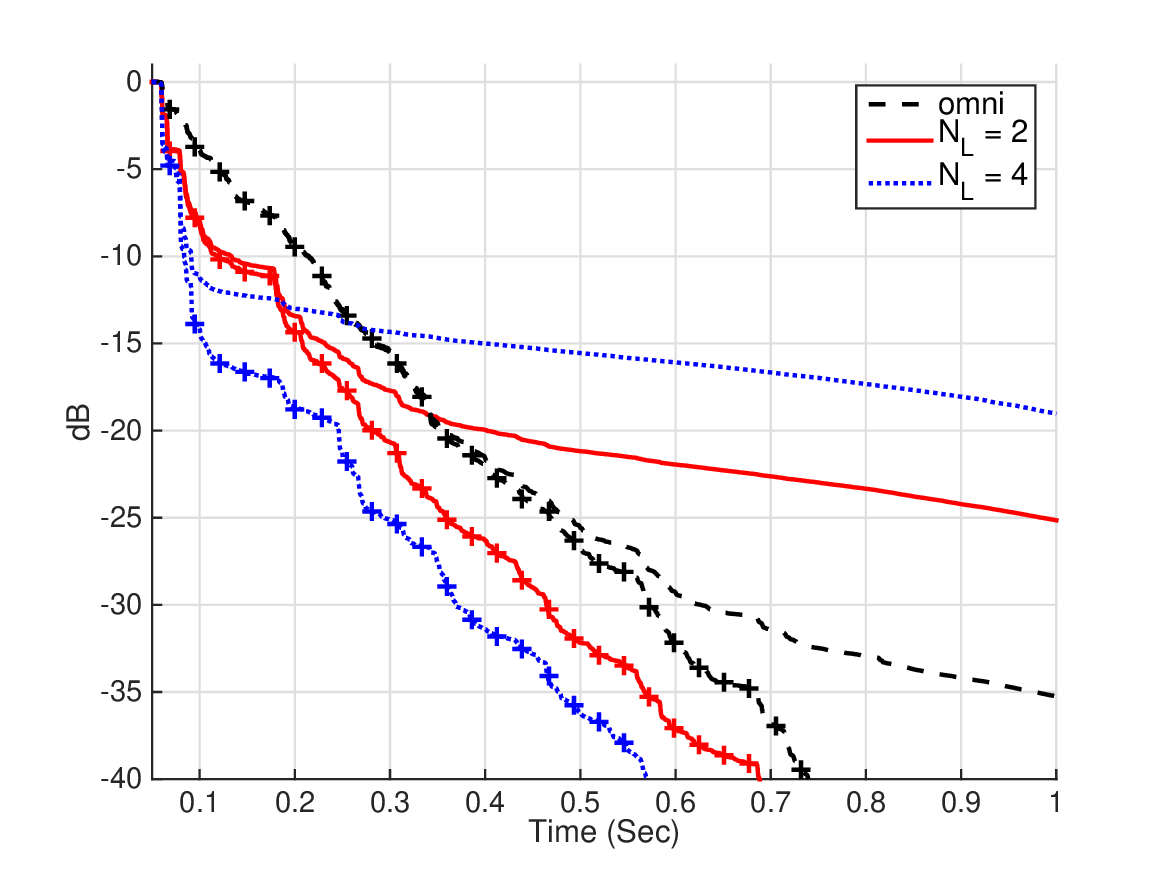}
   \caption{1$\,$kHz}
   \label{fig:Ng2}
\end{subfigure}
\begin{subfigure}[b]{0.49\textwidth}
   \includegraphics[width=0.93\textwidth]{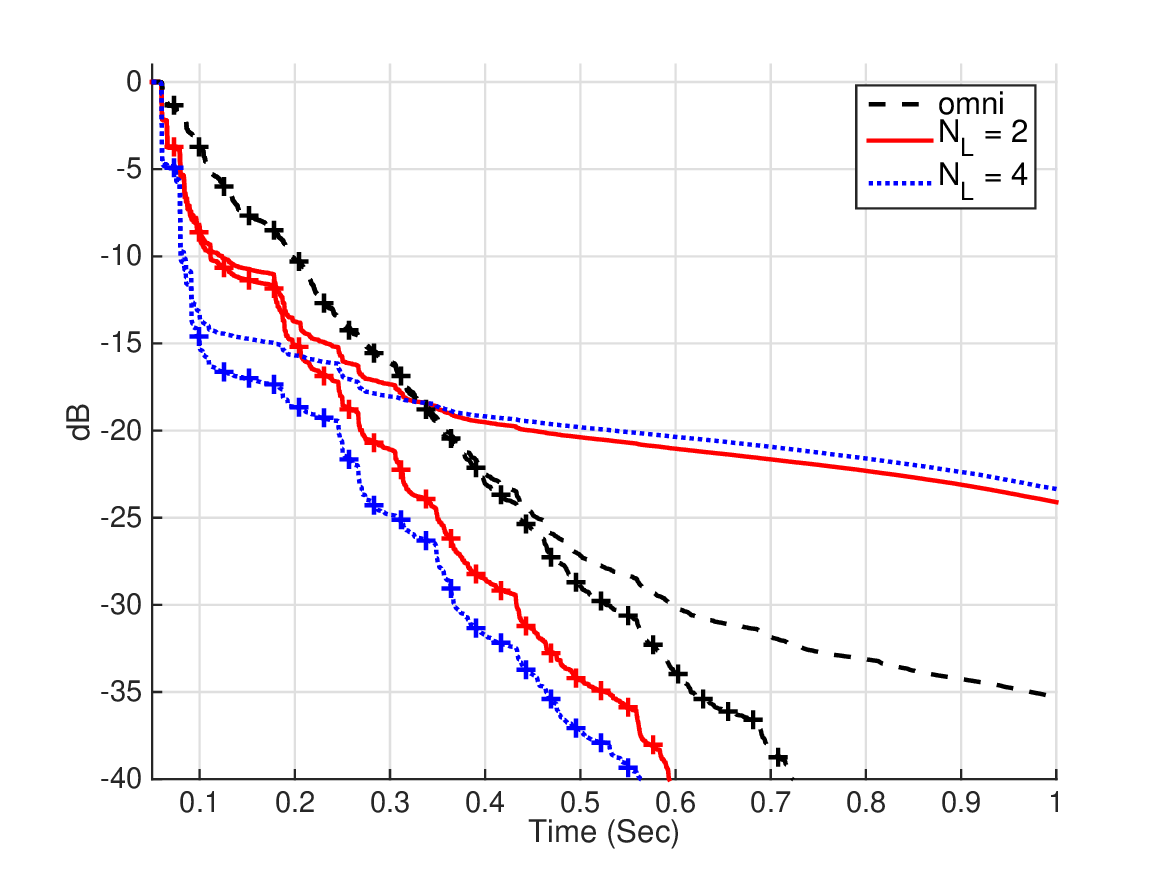}
   \caption{2$\,$kHz}
   \label{fig:Ng2}
\end{subfigure}
\begin{subfigure}[b]{0.49\textwidth}
   \includegraphics[width=0.93\textwidth]{maxC50nor_fcenter4000_SNR-30.eps}
   \caption{4$\,$kHz}
   \label{fig:Ng2}
\end{subfigure}
\caption{EDCs for RIRs with error, leading to an SNR of 30$\,$dB. Analysis in octave bands is presented for $room\,1$ with RT $=1.14\,$s, for $maxC50$, employing SHs order $N_L = 2$ and 4, and for $omni$. 
For reference, EDCs for error-free RIRs are plotted, indicated using `+' signs.}
\label{fig:robust1}
\end{figure}

EDCs in octave bands are presented in Fig.~\ref{fig:robust1}, for $maxC50$, employing SHs order $N_L = 2$ and 4.
In the figure, the average EDCs over 20 realizations of noise, are presented.  
The figure also shows the EDCs for the error-free RIRs for reference, indicated by `+' signs, and the same plots for $omni$. 
In the figure, similar behavior of the EDCs for error-free RIRs is observed for the different octave bands. 
Also, for most curves, the EDCs for RIRs with error have similar behavior to EDCs for error-free RIRs in the early part of the decay curves.
However, at some time instance after the initial decay, where the effect of the error becomes significant, they deviate from the error-free curves.
The figure shows that a design employing $N_L = 2$ is fairly robust to errors at frequencies 1, 2, and 4$\,$kHz. 
Employing $N_L = 4$ shows similar robustness but only at 2 and 4$\,$kHz. 
The more significant deviation from the EDCs for error-free RIRs in octave $0.5\,$kHz, for $N_L = 2$, and octaves 0.5 and $1\,$kHz, for $N_L = 4$, can be explained by the normalization by $\bm B_L^{-1}(k)$ in Eq.~(\ref{eq:model7}). 
Following the analysis in \cite{rafaely2005analysis}, the elements of $\bm B_L(k)$, i.e.,~$b_{n}^L(kr_L)$, have low magnitudes at frequencies for which $n>kr$. 
Therefore, the inversion of $\bm B_L(k)$ typically introduces ill-conditioning for high SHs orders at low frequencies. 
At 1$\,$kHz, $kr_L = 1.8153$, which explains why the system of order $N_L = 4$ is less robust compared to the system of order $N_L = 2$. 

The effect of estimation errors on the clarity of the response is now studied.  
Table~\ref{table:lasteffort} presentes the average $C50$ over 20 realizations of noise for the different octave bands and for $maxC50$ and $minC50$ for $N_L = 2,3,$ and 4, and for $omni$.  
$C50$ values for error-free RIRs are also presented in the table. 
In order to compare these more easily, boldcase font was used for the cases where the $C50$ values for RIRs with error were different by at least $3\,$dB to the values for error-free RIRs. 
\begin{table}
\centering
\small
\begin{tabular}{ | l  l | c | c | c | c | c | c | c | c | }
\hline
\multicolumn{2}{ | c | }{Beaformer and}   & \multicolumn{2}{ | c | }{$0.5\,$kHz} & \multicolumn{2}{ | c | }{$1\,$kHz} & \multicolumn{2}{ | c | }{$2\,$kHz} & \multicolumn{2}{ | c | }{$4\,$kHz}\\
\multicolumn{2}{  | c | }{SLA order} & error-free & error & error-free & error  & error-free & error  & error-free & error\\
\hline 
\multicolumn{2}{  | l | }{ $omni$}  &
3.8 & 3.8 & 2.6 & 2.5 & 3.5 & 3.5 & 3.8 & 3.8\\
\hline
$maxC50$, & $N_L = 2$ & 
9.4 & 7.8 & 9.2 & 8.9 & 9.7 & 9.2 & 10.2 & 9.6\\
 &$N_L = 3$&  
13.8 & \textbf{-1.3} & 12.3 & 11.2 & 12.6 & 11.5 & 13.7 & 12.1\\
&$N_L = 4$ & 
16.4 & \textbf{-12.5} & 15.6 & \textbf{11.7} & 16.2 & 14.1 & 17.1 & 14.6\\
\hline
$minC50$, & $N_L = 2$ & 
-7.4 & -7.9 & -4.6 & -4.9 & -4.7 & -5.0 & -3.8 & -4.2\\
 & $N_L = 3$ &  
-17.3 & \textbf{-13.2} & -12.9 & -13.0 & -13.9 & -13.8 & -12.2 & -12.4\\
 & $N_L = 4$ &
-23.4 & \textbf{-12.5} & -17.4 & -15.1 & -17.2 & -16.2 & -14.9 & -14.5\\
\hline
\end{tabular}
\caption{$C50\,$[dB] for RIRs with error, leading to an SNR of 30$\,$dB, and for error-free RIRs. Analysis in octave bands for $room\,1$ with RT of $1.14\,$s. Values of $C50$ for RIRs with error that are different by at least $3\,$dB compared to those with error-free RIRs are emphasized using boldface font.}
\label{table:lasteffort}
\end{table}
Firstly, the $C50$ values for error-free RIRs in octave bands show similar behavior to those in Table.~\ref{table:clarity2}.
Observing the $C50$ values for RIRs with error, employing $N_L = 2$ is robust to errors at all frequencies, with differences of less than 3$\,$dB in $C50$. 
On the other hand, employing $N_L = 3$ is robust only at 1, 2, and $4\,$kHz, and employing $N_L = 4$ is robust only at $2$ and $4\,$kHz. 
These results are in agreement with the analysis of the EDCs from Fig.~\ref{fig:robust1} for orders $N_L = 2$ and 4. 
Robustness analysis was also performed for higher and lower SNRs, showing similar behavior, i.e., the higher orders at low frequencies become more sensitive to noise as the SNR degrades. 

In conclusion, this section provided a robustness analysis, showing the effects of errors on the EDC and $C50$ values for the proposed beamformers. 
The results show that the beamformers are reasonably robust with limitations imposed by the use of high SHs in low octave bands. 
This motivates the design of beamformers with different SHs orders for different octave bands. 
Furthermore, improving the robustness to errors is proposed for future study, by using, for example, the methods proposed in \cite{jungmann2012combined, lim2014robust} for multiple microphones. 

\section{LISTENING TESTS}\label{sec:listeningtests}

The objective of the listening test is to investigate the effect of the proposed beamformers on perceptual attributes of the sound field surrounding human listeners. 
The investigation was performed by subjectively comparing binaural RIRs synthesized using the proposed beamformers, in order to evaluate the levels of reverberation and sound envelopment. 

\subsection{Methodology}

A speech signal of a male speaker was used as an input signal at the SLA to evaluate the level of reverberation. 
Another signal from a classical guitar was similarly used to evaluate sound envelopment.
The signals were convolved with rendered binaural RIRs, generated using the normalized RTF matrix of $room\,1$ (with RT $1.14\,$s), and a set of pre-measured HRTFs. 
This set was taken from the Cologne HRTF database for the Neumann KU100 dummy head \cite{bernschutz2013spherical}, and was truncated to a SHs order of $N_M = 5$ in the experiment due to computational limitations. 
A binaural representation of the sound field surrounding the SMA was rendered by applying $\bm \lambda^{B}$ with $N_M = 5$ to the SMA (see Eq.~(\ref{eq:model6b})), for the left and right ears, and using SHs coefficients of the respective HRTF set. 
\textcolor{hai}{
In particular, since $\bm \lambda^{B}$ involves coefficients that vary over frequency, beamforming at the SMA is applied in the frequency domain. 
Then, the resulting vector is transformed to the time domain, using the inverse DFT.}
The SLA beamformers from Sec.~\ref{sec:simulationmethods}, were then applied to the SLA, employing a SHs order $N_L = 4$ and, in addition, $\bm \gamma^O$ was set for modelling an omnidirectional source (as described in Sec.~\ref{sec:simulationresults}). 
With the aim of including horizontal head-tracking (which is required for achieving an effective spatial realism), binaural responses were computed for each head rotation (spanning $360^\circ$ with a 1$^\circ$ resolution) by rotating the HRTFs. 
Rotation was performed by multiplying the SHs representation of the HRTF set by Wigner-D functions \cite{rafaely2008spherical2}. 
For sound reproduction, a pair of AKG-K701 headphones were fitted with a Razor IMU sensor for head tracking, which is a nine degrees of freedom Attitude and Heading Reference System (AHRS).
All stimuli were processed with a matching headphone compensation filter, were generated pre-test, and were played-back using the SoundScape Renderer auralization engine \cite{ahrens2008soundscape}.

Fourteen normal hearing subjects (10 male, 4 female, ages 20-52) participated in a \textcolor{hai}{MUSHRA-like (multiple stimuli with hidden reference and anchor) listening experiment \cite{recommendation2003method}.}
The test details and the differences from a full MUSHRA test are provided below. 
The experiment included two separate tests, and for both tests the labeled reference was based on a binaural RIR that corresponds to $minC50$. 
In the first test, temporal attributes of the sound fields were investigated; 
listeners were asked to rank the reverberation level of five responses of the speech signal, corresponding to the five SLA beamformers, and relative to a reference response. 
In the second test, the listeners were required to rank the sound envelopment level of the five responses of the music signal, corresponding to the same beamformers, relative to a reference response. 
This test investigates the spatial attributes of the sound field. 
A ranking scale of 0-100 was used, and listeners were required to rank the responses that have the most similar levels of reverberation or envelopment as the respective reference signals with a score of 100. 
The other signals were ranked accordingly. 
\textcolor{hai}{Note that in this test the reference signal was chosen as the $minC50$ signal, and no anchor was used. 
Although this is not standard practice in MUSHRA tests, it avoided the use of an inappropriate reference and anchor that may have widened the ranking range, therefore unnecessarily reducing the differences between test signals. 
The development of reliable reference and anchor signals for this test is proposed for future work. 
For this reason the test is referred to as a MUSHRA-like.}
Each human listener performed each test twice, and results were averaged.
Finally, the results of one listener were omitted since the listener was identified as an outlier in the post-screening of the subjects \cite{recommendation2003method}. 

\subsection{Results}

\begin{figure}
\centering
\includegraphics[width=23pc]{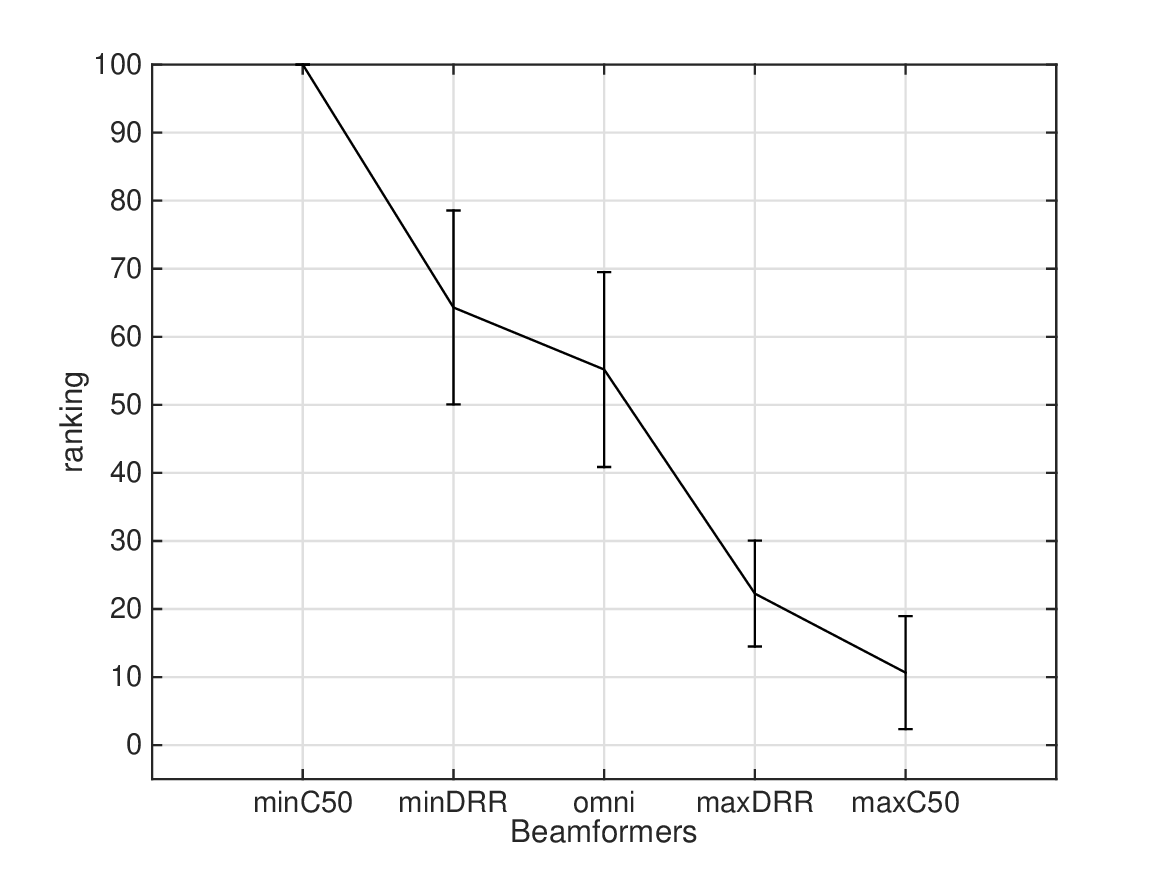}
\caption{Mean scores and 95$\%$ confidence intervals for the subjective evaluation of the reverberation level for the five SLA beamformers.}
\label{fig:listeningrev}
\end{figure}

\begin{figure}
\centering
\includegraphics[width=23pc]{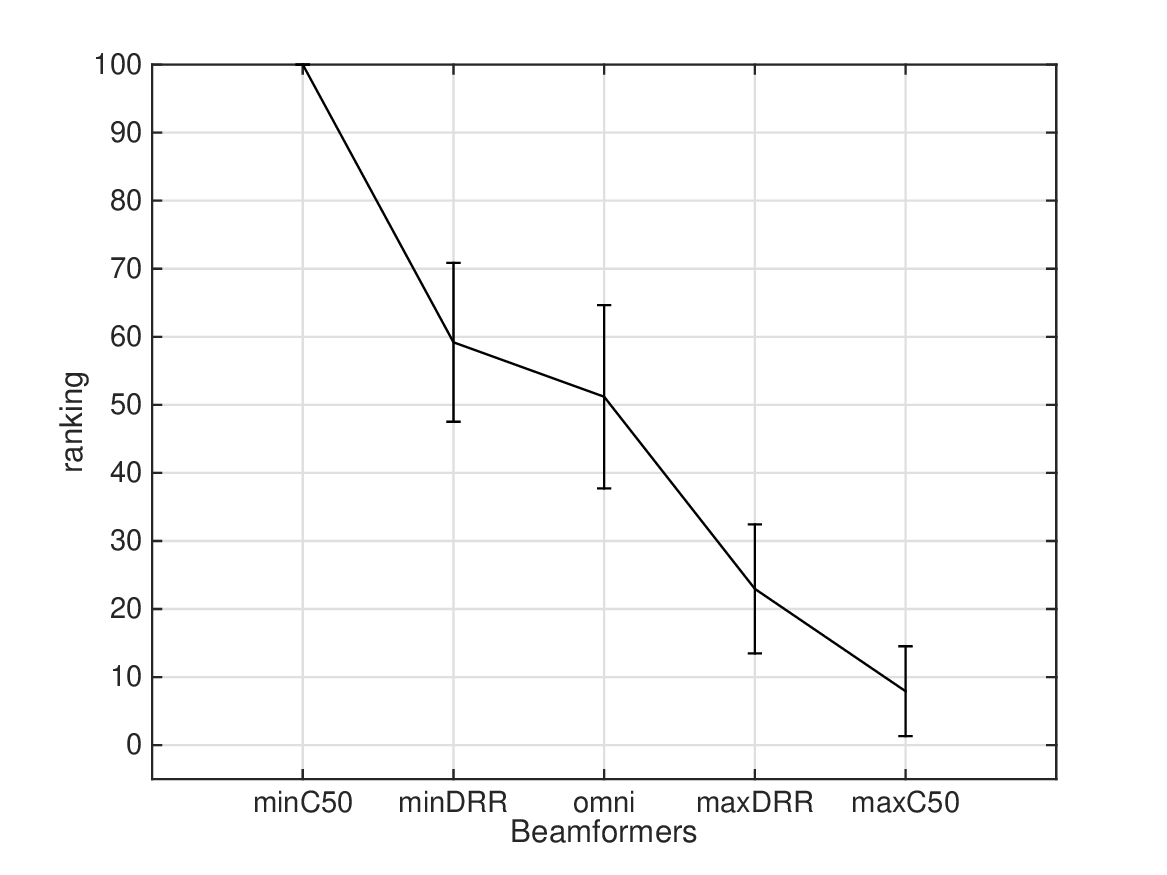}
\caption{Mean scores and 95$\%$ confidence intervals for the subjective evaluation of the sound envelopment level for the five SLA beamformers.}
\label{fig:listeningsound}
\end{figure}

Figures \ref{fig:listeningrev} and \ref{fig:listeningsound} show the mean scores and 95$\%$ confidence intervals ($t_{.95,13} = 2.06$) for the five SLA beamformers in the first and second experiments, respectively. 
The results demonstrate that applying beamforming to an SLA with the proposed methods can lead to perceivable changes in the sound field surrounding human listeners in a room. 
This validates the theoretical and simulation results presented in the previous sections.
$minC50$ was chosen as the reference and, therefore, received a score of 100 for the levels of both reverberation and sound envelopment. 
The mean scores for the rest of the beamformers are arranged in the same descending order in both tests; $minDRR$, $omni$, $maxDRR$, and $maxC50$. 
$minDRR$ and $omni$ also received relatively high scores for reverberation and sound envelopment. 
Due to a significant overlapping in their 95$\%$ confidence intervals, applying these beamformers is expected to have a similar effect on human listeners. 
Finally, $maxDRR$ and $maxC50$ received relatively low scores in both tests, with smaller 95$\%$ confidence intervals, compared to those of $minDRR$ and $omni$.  
In particular, $maxC50$ shows the lowest levels of reverberation and sound envelopment and is, therefore, regarded as the beamformer of choice for spatial dereverberation within the tested set.
To summarize, the significant differences between omni and both $minC50$ and $maxC50$ suggest that the proposed methods generate perceivable reverberation and dereverberation. 

\section{CONCLUSIONS}\label{sec:conclusions}

In this paper, beamforming methods were proposed for modifying the reverberation attributes of a sound field around a listener using compact SLAs. 
It has been shown that employing these methods changes both the temporal and the spatial attributes of a sound field. 
The robustness of these methods to RIR estimation errors was investigated, showing good robustness for specific combinations of the SLA SHs order and a chosen octave band. 
Employment of the methods was shown to result in perceivable differences in hearing tests with simulated RIRs.
The results imply that sound systems with directional sources may be employed for spatial dereverberation and reverberation. 
Extending the proposed methods for several listening zones, instead of a single listener position, as well as an experimental validation are proposed for future work. 

\section{ACKNOWLEDGMENT}
This investigation was supported by The Israel Science Foundation (Grant No.~146/13). 
The authors would like to thank Zamir Ben-Hur for his technical assistance with the listening tests. 

\bibliographystyle{IEEEtran}
\bibliography{allbib2}

\begin{thebibliography}{10}
\providecommand{\url}[1]{#1}
\csname url@samestyle\endcsname
\providecommand{\newblock}{\relax}
\providecommand{\bibinfo}[2]{#2}
\providecommand{\BIBentrySTDinterwordspacing}{\spaceskip=0pt\relax}
\providecommand{\BIBentryALTinterwordstretchfactor}{4}
\providecommand{\BIBentryALTinterwordspacing}{\spaceskip=\fontdimen2\font plus
\BIBentryALTinterwordstretchfactor\fontdimen3\font minus \fontdimen4\font\relax}
\providecommand{\BIBforeignlanguage}[2]{{%
\expandafter\ifx\csname l@#1\endcsname\relax
\typeout{** WARNING: IEEEtran.bst: No hyphenation pattern has been}%
\typeout{** loaded for the language `#1'. Using the pattern for}%
\typeout{** the default language instead.}%
\else
\language=\csname l@#1\endcsname
\fi
#2}}
\providecommand{\BIBdecl}{\relax}
\BIBdecl

\bibitem{beranek2012concert}
L.~Beranek, \emph{Concert halls and opera houses: music, acoustics, and architecture}.\hskip 1em plus 0.5em minus 0.4em\relax Springer Science \& Business Media, 2012.

\bibitem{bistafa2000reverberation}
S.~R. Bistafa and J.~S. Bradley, ``Reverberation time and maximum background-noise level for classrooms from a comparative study of speech intelligibility metrics,'' \emph{The Journal of the Acoustical Society of America}, vol. 107, no.~2, pp. 861--875, 2000.

\bibitem{kinoshita2013reverb}
K.~Kinoshita, M.~Delcroix, T.~Yoshioka, T.~Nakatani, A.~Sehr, W.~Kellermann, and R.~Maas, ``The reverb challenge: Acommon evaluation framework for dereverberation and recognition of reverberant speech,'' in \emph{Applications of Signal Processing to Audio and Acoustics (WASPAA), 2013 IEEE Workshop on}.\hskip 1em plus 0.5em minus 0.4em\relax IEEE, 2013, pp. 1--4.

\bibitem{gillespie2002acoustic}
B.~W. Gillespie and L.~E. Atlas, ``Acoustic diversity for improved speech recognition in reverberant environments,'' in \emph{Acoustics, Speech, and Signal Processing (ICASSP), 2002 IEEE International Conference on}, vol.~1.\hskip 1em plus 0.5em minus 0.4em\relax IEEE, 2002, pp. I--557.

\bibitem{gustafsson2003source}
T.~Gustafsson, B.~D. Rao, and M.~Trivedi, ``Source localization in reverberant environments: Modeling and statistical analysis,'' \emph{Speech and Audio Processing, IEEE Transactions on}, vol.~11, no.~6, pp. 791--803, 2003.

\bibitem{neely1979invertibility}
S.~T. Neely and J.~B. Allen, ``Invertibility of a room impulse response,'' \emph{The Journal of the Acoustical Society of America}, vol.~66, no.~1, pp. 165--169, 1979.

\bibitem{kallinger2005impulse}
M.~Kallinger and A.~Mertins, ``Impulse response shortening for acoustic listening room compensation,'' in \emph{Proc. International Workshop on Acoustic Echo and Noise Control (IWAENC)}, 2005, pp. 197--200.

\bibitem{mertins2010room}
A.~Mertins, T.~Mei, and M.~Kallinger, ``Room impulse response shortening/reshaping with infinity-and-norm optimization,'' \emph{IEEE Transactions on Audio, Speech, and Language Processing}, vol.~18, no.~2, pp. 249--259, 2010.

\bibitem{standard19973382}
I.~Standard, ``3382. acoustics--measurement of the reverberation time of rooms with reference to other acoustical parameters,'' \emph{International Standards Organization}, 1997.

\bibitem{miyoshi1988inverse}
M.~Miyoshi and Y.~Kaneda, ``Inverse filtering of room acoustics,'' \emph{IEEE Transactions on Acoustics, Speech, and Signal Processing}, vol.~36, no.~2, pp. 145--152, 1988.

\bibitem{zhang2010use}
W.~Zhang, E.~A. Habets, and P.~A. Naylor, ``On the use of channel shortening in multichannel acoustic system equalization,'' in \emph{Proc. Int. Workshop Acoust. Echo Noise Control (IWAENC)}, 2010.

\bibitem{lim2014robust}
F.~Lim, W.~Zhang, E.~A. Habets, and P.~A. Naylor, ``Robust multichannel dereverberation using relaxed multichannel least squares,'' \emph{IEEE/ACM Transactions on Audio, Speech, and Language Processing}, vol.~22, no.~9, pp. 1379--1390, 2014.

\bibitem{mei2010robustness}
T.~Mei and A.~Mertins, ``On the robustness of room impulse response reshaping,'' in \emph{Proc. International Workshop on Acoustic Echo and Noise Control (IWAENC 2010), Tel Aviv, Israel}, 2010.

\bibitem{jungmann2012combined}
J.~O. Jungmann, R.~Mazur, M.~Kallinger, T.~Mei, and A.~Mertins, ``Combined acoustic mimo channel crosstalk cancellation and room impulse response reshaping,'' \emph{IEEE Transactions on Audio, Speech, and Language Processing}, vol.~20, no.~6, pp. 1829--1842, 2012.

\bibitem{pollow2009variable}
M.~Pollow and G.~K. Behler, ``Variable directivity for platonic sound sources based on spherical harmonics optimization,'' \emph{Acta Acustica united with Acustica}, vol.~95, no.~6, pp. 1082--1092, 2009.

\bibitem{warusfel2001directivity}
O.~Warusfel and N.~Misdariis, ``Directivity synthesis with a 3d array of loudspeakers: application for stage performance,'' in \emph{Proceedings of the COST G-6 Conference on Digital Audio Effects (DAFX-01), Limerick, Ireland}, 2001, pp. 1--5.

\bibitem{rafaely2009spherical}
B.~Rafaely, ``Spherical loudspeaker array for local active control of sound,'' \emph{The Journal of the Acoustical Society of America}, vol. 125, p. 3006, 2009.

\bibitem{morgenstern2015theory}
H.~Morgenstern, B.~Rafaely, and F.~Zotter, ``Theory and investigation of acoustic multiple-input multiple-output systems based on spherical arrays in a room,'' \emph{The Journal of the Acoustical Society of America}, vol. 138, no.~5, pp. 2998--3009, 2015.

\bibitem{kassakian2003design}
P.~Kassakian and D.~Wessel, ``Design of low-order filters for radiation synthesis,'' in \emph{Audio Engineering Society Convention 115}.\hskip 1em plus 0.5em minus 0.4em\relax Audio Engineering Society, 2003.

\bibitem{zotter2007modeling}
F.~Zotter, A.~Sontacchi, and R.~Holdrich, ``Modeling a spherical loudspeaker system as multipole source,'' \emph{Fortschritte der Akustik}, vol.~33, no.~1, p. 221, 2007.

\bibitem{gerzon1975recording}
M.~A. Gerzon, ``Recording concert hall acoustics for posterity,'' \emph{Journal of the Audio Engineering Society}, vol.~23, no.~7, pp. 569--571, 1975.

\bibitem{farina2006room}
A.~Farina, ``Room impulse responses as temporal and spatial filters,'' in \emph{The 9th Western Pacific Acoustics Conference, Seoul, Korea}, 2006.

\bibitem{morgenstern2012joint}
H.~Morgenstern, F.~Zotter, and B.~Rafaely, ``Joint spherical beam forming for directional analysis of reflections in rooms,'' \emph{The Journal of the Acoustical Society of America}, vol. 131, no.~4, pp. 3207--3207, 2012.

\bibitem{morgenstern2013enhanced}
H.~Morgenstern and B.~Rafaely, ``Enhanced spatial analysis of room acoustics using acoustic multiple-input multiple-output (mimo) systems,'' in \emph{Proceedings of Meetings on Acoustics}, vol.~19, no.~1.\hskip 1em plus 0.5em minus 0.4em\relax Acoustical Society of America, 2013, p. 015018.

\bibitem{pasqual2010application}
A.~M. Pasqual, A.~de~Fran{\c{c}}a, J.~Roberto, and P.~Herzog, ``Application of acoustic radiation modes in the directivity control by a spherical loudspeaker array,'' \emph{Acta acustica united with Acustica}, vol.~96, no.~1, pp. 32--42, 2010.

\bibitem{morgenstern2013sound}
H.~Morgenstern, N.~Shabtai, and B.~Rafaely, ``Sound-field control in enclosures by spherical arrays,'' in \emph{Audio Engineering Society Conference: 52nd International Conference: Sound Field Control-Engineering and Perception}.\hskip 1em plus 0.5em minus 0.4em\relax Audio Engineering Society, 2013.

\bibitem{rafaely2011optimal}
B.~Rafaely and D.~Khaykin, ``Optimal model-based beamforming and independent steering for spherical loudspeaker arrays,'' \emph{Audio, Speech, and Language Processing, IEEE Transactions on}, vol.~19, no.~7, pp. 2234--2238, 2011.

\bibitem{park2005sound}
M.~Park and B.~Rafaely, ``Sound-field analysis by plane-wave decomposition using spherical microphone array,'' \emph{The Journal of the Acoustical Society of America}, vol. 118, no.~5, pp. 3094--3103, 2005.

\bibitem{morgenstern2014farfield}
H.~Morgenstern and B.~Rafaely, ``Far-field criterion for spherical microphone arrays and directional sources,'' in \emph{Hands-free Speech Communication and Microphone Arrays (HSCMA), 2014 Joint Workshop on}, 2014.

\bibitem{rafaely2004plane}
B.~Rafaely, ``Plane-wave decomposition of the sound field on a sphere by spherical convolution,'' \emph{The Journal of the Acoustical Society of America}, vol. 116, p. 2149, 2004.

\bibitem{williams1999fourier}
E.~Williams, \emph{Fourier acoustics: sound radiation and nearfield acoustical holography}.\hskip 1em plus 0.5em minus 0.4em\relax Academic Press, 1999.

\bibitem{tourbabin2015consistent}
V.~Tourbabin and B.~Rafaely, ``On the consistent use of space and time conventions in array processing,'' \emph{Acta Acustica united with Acustica}, vol. 101, no.~3, pp. 470--473, 2015.

\bibitem{allen1979image}
J.~B. Allen and D.~A. Berkley, ``Image method for efficiently simulating small-room acoustics,'' \emph{The Journal of the Acoustical Society of America}, vol.~65, p. 943, 1979.

\bibitem{li2007flexible}
Z.~Li and R.~Duraiswami, ``Flexible and optimal design of spherical microphone arrays for beamforming,'' \emph{Audio, Speech, and Language Processing, IEEE Transactions on}, vol.~15, no.~2, pp. 702--714, 2007.

\bibitem{rafaely2010interaural}
B.~Rafaely and A.~Avni, ``Interaural cross correlation in a sound field represented by spherical harmonics,'' \emph{The Journal of the Acoustical Society of America}, vol. 127, no.~2, pp. 823--828, 2010.

\bibitem{kinsler1999fundamentals}
L.~E. Kinsler, A.~R. Frey, A.~B. Coppens, and J.~V. Sanders, ``Fundamentals of acoustics,'' \emph{Fundamentals of Acoustics, 4th Edition, by Lawrence E. Kinsler, Austin R. Frey, Alan B. Coppens, James V. Sanders, pp. 560. ISBN 0-471-84789-5. Wiley-VCH, December 1999.}, vol.~1, 1999.

\bibitem{naylor2010speech}
P.~A. Naylor and N.~D. Gaubitch, \emph{Speech dereverberation}.\hskip 1em plus 0.5em minus 0.4em\relax Springer Science \& Business Media, 2010.

\bibitem{avizienis2006compact}
R.~Avizienis, A.~Freed, P.~Kassakian, and D.~Wessel, ``A compact 120 independent element spherical loudspeaker array with programable radiation patterns,'' in \emph{Audio Engineering Society Convention 120}.\hskip 1em plus 0.5em minus 0.4em\relax Audio Engineering Society, 2006.

\bibitem{meyer2002highly}
J.~Meyer and G.~Elko, ``A highly scalable spherical microphone array based on an orthonormal decomposition of the soundfield,'' in \emph{Acoustics, Speech, and Signal Processing (ICASSP), 2002 IEEE International Conference on}, vol.~2.\hskip 1em plus 0.5em minus 0.4em\relax IEEE, 2002, pp. II--1781.

\bibitem{wabnitz2010room}
A.~Wabnitz, N.~Epain, C.~Jin, and A.~van Schaik, ``Room acoustics simulation for multichannel microphone arrays,'' in \emph{Proceedings of the International Symposium on Room Acoustics}, 2010.

\bibitem{morgenstern2015joint}
H.~Morgenstern, B.~Rafaely, and M.~Noisternig, ``Joint design of spherical microphone and loudspeaker arrays for room acoustic analysis,'' in \emph{41th Annual German Congress on Acoustics (DAGA)}, vol.~41, 2015.

\bibitem{van2004detection}
H.~L. Van~Trees, \emph{Detection, Estimation, and Modulation Theory, Optimum Array Processing}.\hskip 1em plus 0.5em minus 0.4em\relax John Wiley \& Sons, 2004.

\bibitem{elko2004differential}
G.~W. Elko, ``Differential microphone arrays,'' in \emph{Audio signal processing for next-generation multimedia communication systems}.\hskip 1em plus 0.5em minus 0.4em\relax Springer, 2004, pp. 11--65.

\bibitem{schroeder1965new}
M.~R. Schroeder, ``New method of measuring reverberation time,'' \emph{The Journal of the Acoustical Society of America}, vol.~37, no.~3, pp. 409--412, 1965.

\bibitem{rafaely2005analysis}
B.~Rafaely, ``Analysis and design of spherical microphone arrays,'' \emph{Speech and Audio Processing, IEEE Transactions on}, vol.~13, no.~1, pp. 135--143, 2005.

\bibitem{bernschutz2013spherical}
B.~Bernsch{\"u}tz, ``A spherical far field hrir/hrtf compilation of the neumann ku 100,'' in \emph{Proceedings of the 40th Italian (AIA) Annual Conference on Acoustics and the 39th German Annual Conference on Acoustics (DAGA) Conference on Acoustics}, 2013, p.~29.

\bibitem{rafaely2008spherical2}
B.~Rafaely and M.~Kleider, ``Spherical microphone array beam steering using wigner-d weighting,'' \emph{Signal Processing Letters, IEEE}, vol.~15, pp. 417--420, 2008.

\bibitem{ahrens2008soundscape}
J.~Ahrens, M.~Geier, and S.~Spors, ``The soundscape renderer: A unified spatial audio reproduction framework for arbitrary rendering methods,'' in \emph{Audio Engineering Society Convention 124}.\hskip 1em plus 0.5em minus 0.4em\relax Audio Engineering Society, 2008.

\bibitem{recommendation2003method}
I.~Recommendation, ``Method for the subjective assessment of intermediate quality level of coding systems,'' \emph{ITU-R BS}, pp. 1534--1, 2003.

\end{thebibliography}

\end{document}